\documentclass[journal]{IEEEtran}
\usepackage{amsmath,amssymb,amsfonts}
\usepackage[caption=false,font=normalsize,labelfont=sf,textfont=sf]{subfig}
\usepackage{stfloats}
\usepackage{verbatim}
\usepackage{graphicx}
\usepackage{bm}
\usepackage{cite}
\usepackage{color}
\usepackage[]{hyperref}

\usepackage[linesnumbered, ruled]{algorithm2e}
\SetKwRepeat{Do}{do}{while}

\begin{document}

\title{UAV-Assisted Integrated Communication and Over-the-Air Computation with \\Interference Awareness}
\author{Xunqiang Lan,
		Xiao Tang,
        Ruonan Zhang,
        Bin Li,
        Yichen Wang,
        Dusit Niyato,
        and Zhu Han
\thanks{X. Lan, R. Zhang, and B. Li are with the School of Electronics and Information, Northwestern Polytechnical University, Xi'an 710072, China. (email: lanxunqiang@mail.nwpu.edu.cn)}
\thanks{X. Tang and Y. Wang are with the School of Information and Communication Engineering, Xi'an Jiaotong University, Xi'an 710049, China.}
\thanks{D. Niyato is with the School of Computer Science and Engineering, Nanyang Technological University, Singapore.}
\thanks{Z. Han is with the Department of Electrical and Computer Engineering at the University of Houston, Houston, TX 77004 USA, and also with the Department of Computer Science and Engineering, Kyung Hee University, Seoul 446-701, South Korea.}
}

\maketitle

\begin{abstract}
Over-the-air computation (AirComp) is a promising technique that addresses big data collection and fast wireless data aggregation. However, in a network where wireless communication and AirComp coexist, mutual interference becomes a critical challenge. In this paper, we propose to employ an unmanned aerial vehicle (UAV) to enable integrated communication and AirComp, where we capitalize on UAV mobility with alleviated interference for performance enhancement. Particularly, we aim to maximize the sum of user transmission rate with the guaranteed AirComp accuracy requirement, where we jointly optimize the transmission strategy, signal normalizing factor, scheduling strategy, and UAV trajectory. We decouple the formulated problem into two layers where the outer layer is for UAV trajectory and scheduling, and the inner layer is for transmission and computation. Then, we solve the inner layer problem through alternating optimization, and the outer layer is solved through soft actor–critic-based deep reinforcement learning. Simulation results show the convergence of the proposed learning process and also demonstrate the performance superiority of our proposal as compared with the baselines in various situations.
\end{abstract}


\section{Introduction}

\IEEEPARstart{T}{he} future 6G networks are expected to empower the automation, informatization, and intelligence of applications, thereby driving an evolution toward a connected intelligent world~\cite{6G}. Toward this vision, a wide array of smart sensors/devices have been deployed, integrating sensing, communication, and computation, and establishing the foundation for numerous data-intensive applications. However, providing the spectrum and computing resources necessary for massive wireless connectivity and data computation for such a large number of devices may overwhelm network capacity~\cite{spectrum}. On the other hand, certain categories of applications typically require functional computations on distributed data to enable rapid collection and aggregation of large datasets for final decision-making, rather than focusing on reconstructing individual data. Examples include updating artificial intelligence models in distributed learning~\cite{app1} and monitoring temperature, humidity, and chemical levels in sensing~\cite{app2,app3}. Therefore, achieving rapid aggregation of massive data without relying on the reconstruction of individual data is an important research direction, as it can reduce communication overhead and latency, thus enhancing the system's efficiency~\cite{latency}.

Aiming at the issue above, over-the-air computation (AirComp) has emerged as a promising approach. By leveraging the inherent broadcast nature of wireless communications and the application of mathematical functions, AirComp can effectively aggregate sensor data from a large number of concurrent sensor transmissions~\cite{aircomp}. Specifically, AirComp exploits the waveform/signal superposition property of multiple access channels and the functional decomposition, where all signals from concurrent devices can inherently execute an additive operation over the same radio channel. The fusion center can then obtain a nomographic function of the distributed data (e.g., mean and geometric mean) from these concurrent sensor transmissions. With advantages such as large-scale device access, low latency, high spectral efficiency, and minimal computational overhead, AirComp has garnered significant attention for data-intensive and distributed data aggregation applications~\cite{advan}.

However, the performance of AirComp is significantly influenced by network deployment and is constrained by transmission channel quality. Data transmitted by sensor devices with poor channel quality can reduce the reliability and accuracy of overall data aggregation~\cite{csi}. Challenged by the precise and reliable data transmission requirements of AirComp, it is crucial to maintain high-quality transmission channels for the devices. In this regard, unmanned aerial vehicle (UAV)-assisted AirComp has emerged as a highly promising alternative. Attracted by the wide and flexible applications of UAVs, we can deploy UAVs as aerial base stations to reach the vicinity of devices that can establish dynamically flexible, efficient, and stable communication networks, especially in cases of traditional infrastructure failures or coverage gaps~\cite{uav1}. Supported by UAV, AirComp's sensor devices can easily extend to network edge areas, remote regions, or harsh environments without the need to build traditional ground communication infrastructure. Furthermore, due to the high deployment altitude of UAVs, typically with line-of-sight (LoS) connectivity to ground devices, the probability of deep fading in channels is reduced (typical Rician channels), enhancing signal alignment. In this regard, UAV-assisted AirComp not only improves the performance of AirComp but also reduces the cost~\cite{uav2}.

Owing to the benefits of UAV-assisted communication and AirComp, the application scenarios integrating UAV-assisted communication with UAV-assisted AirComp have attracted significant attention~\cite{inter1},~\cite{inter2}. However, this integration not only brings improvements in communication efficiency and more effective utilization of spectrum resources but also introduces new challenges, particularly the inevitable interference between communication and AirComp. 
The interference is particularly sensitive for UAV aerial base stations, which are limited in both energy and computational power. In addition, the result of AirComp for distributed sensing data is typically used for event inference, which can be accurately performed once the computation error falls below a certain threshold. Therefore, it is essential to consider interactions of transmission strategy, user scheduling strategy, and UAV trajectory planning when studying the use of UAVs in integrated communication and AirComp scenarios.
This approach can effectively alleviate the interference between communication and AirComp, maintain AirComp's computational accuracy, and increase the number of tasks executed within a given time frame. However, most existing work on UAV-assisted AirComp primarily focuses on optimizing computational accuracy strategies within single or multiple AirComp network scenarios, neglecting the {\em analysis of communication and computation requirements} and the {\em interference issues} in the complex scenarios of UAV-assisted integrated communication and AirComp.

To address the aforementioned issues, in this paper, we aim to maximize the user transmission rate for UAV-assisted communication and AirComp while considering the constraint of AirComp accuracy. In particular, we jointly optimize transmission strategy, scheduling strategy, and UAV trajectory planning in UAV-assisted integrated communication and AirComp network to manage interference between users and AirComp. The main contributions can be summarized as follows:
\begin{itemize}
\item We innovatively propose an integrated communication and AirComp framework, where a UAV is deployed to perform data aggregation for AirComp while simultaneously providing communication to ground users. In this regard, we model the communication and AirComp associated with UAV flying way-points in the presence of mutual interference.
\item We formulate the problem to maximize the sum of user communication rates under the constraint of AirComp computational mean-square error (MSE). Moreover, we jointly investigate ground device transmission, scheduling strategies, and UAV trajectory design to alleviate the mutual interference between communication and AirComp efficiently.
\item The problem is decomposed into two layers, one addressing the inner layer of transmission and normalization factor, and the other addressing the outer layer of user scheduling and UAV trajectory optimization. The inner and outer layers are then solved separately using an iterative algorithm and the proposed soft actor-critic (SAC)-based deep reinforcement learning (DRL) algorithm.
\item Simulation results indicate that the considered user communication rate decreases as the AirComp accuracy increases, revealing a trade-off between user communication performance and AirComp accuracy in the model with mutual interference between communication and AirComp. Furthermore, the simulation results demonstrate the effectiveness and superiority of the designed SAC-based algorithm.
\end{itemize}

The rest of this paper is organized as follows. In Sec.~\ref{sec:rw}, we review the related works. In Sec.~\ref{sec:sys}, we introduce the system model of UAV-assisted integrated communication, as well as the formulation and analysis of the problem. In Sec.~\ref{sec:inner}, the inner problem is solved for the transmission strategy and normalization factor optimization. In Sec.~\ref{sec:outer}, a SAC-based algorithm for solving the user scheduling and UAV trajectory is introduced. Sec.~\ref{sec:sim} provides the simulation results to demonstrate the performance, and finally, Sec.~\ref{sec:con} concludes this paper.
\vspace{-2mm}

\section{Related Works} \label{sec:rw}
\subsection{AirComp}
Due to the capability of fast wireless data aggregation, AirComp has recently attracted research interests in various topics~\cite{ac0}. In~\cite{ac1}, the authors proposed a task-oriented integrated sensing, computation, and communication scheme with AirComp to enhance the utilization of limited network resources in edge-device co-inference tasks. In~\cite{ac2}, the authors proposed an over-the-air federated edge learning solution to improve communication efficiency by controlling transmission power to combat aggregation errors. In~\cite{ac3}, the authors designed a multiple-input-multiple output AirComp equalization and channel feedback techniques for spatially multiplexing multifunction computation to design the orthogonal feedback channels. In~\cite{ac4}, the authors considered that the performance of AirComp is constrained by the worst channel conditions and propose a reconfigurable intelligent surface (RIS) assisted AirComp system, which boosts the received signal power through a joint design of AirComp transceivers and RIS phase shift. In~\cite{ac5}, the authors combined AirComp and energy beamforming to design wireless-powered AirComp while using RIS to enhance the efficiency of links and minimize the MSE of AirComp distortion. In~\cite{ac6}, the authors addressed the poor channel conditions between the fusion center and distributed wireless devices and adopted RISs to minimize the average computational MSE of AirComp. Moreover, the Fluid Antenna Systems can dynamically reconfigure position and radiation pattern and also have been employed in integrated sensing and communication architectures to significantly enhance sensing accuracy and communication throughput~\cite{fas1,fas2}.
Existing AirComp researches above are based on static terrestrial fusion centers and single-application scenarios, which struggle to address the dual challenges of limited coverage in remote areas (e.g., mountainous or desert regions) and complex radio environments. To fully exploit the advantages of AirComp, it is imperative to deploy flexible fusion centers and incorporate practical complex network scenarios into the optimization framework.
\begin{figure*}[t]
	\centering
	\includegraphics[width=18cm]{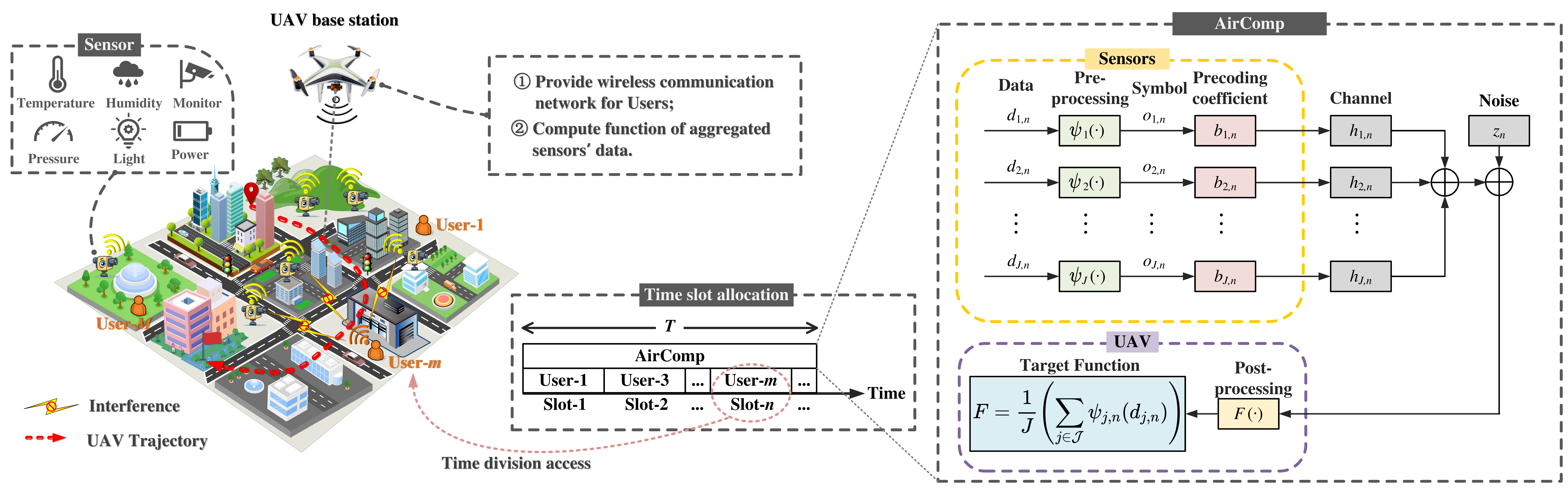}
	\caption{System model.}
	\vspace{-1.5mm}
	\label{fig:sys}
\end{figure*}

\subsection{UAV-Assisted AirComp Network}
UAV-facilitated wireless communication naturally appears as a flexible and effective solution for AirComp, and therefore attracted research interests in UAV-assisted AirComp networks~\cite{uac0}. In~\cite{uac1}, the authors investigated a power-limited multi-slot UAV-assisted AirComp system, which jointly optimizes resource and trajectory to minimize the MSE of AirComp calculation. In~\cite{uac2}, the authors considered a UAV-aided AirComp system, jointly optimizing trajectory, receiver normalization factors, and transmit power of sensors, and proposed a novel equivalent problem transformation to minimize the MSE of AirComp. Similarly, in~\cite{uac3}, the authors studied the interference management in multi-cluster UAVs-assisted AirComp networks to maximize the minimum amount of AirComp tasks executed by each cluster while considering the accuracy requirement. In~\cite{uac4}, the authors investigated the impact of imperfect channel state information on the MSE performance of AirComp, demonstrating the significance of accurate channel estimation for AirComp. In~\cite{uac5}, the authors proposed a non-coherent UAV-assisted AirComp scheme that modulates the amplitude of complementary sequence elements based on the symbols of the parameters to be aggregated, to reliably calculate majority voting in fading channels. In~\cite{uac6}, the authors analyzed a data collection system for multi-UAVs and propose an efficient AirComp strategy using a space timeline code scheme to achieve full space diversity gain, significantly reducing the overall operation time of data collection and the MSE of AirComp. The aforementioned work demonstrates that UAVs can serve as flexible and powerful agents to enhance AirComp performance in various aspects. 
However, these studies primarily involve a UAV serving as an aerial aggregation center, focusing on optimizing communication resources and UAV deployment or trajectory to achieve various AirComp-related objectives, such as maximizing computation accuracy and energy efficiency~\cite{uav01, uav02, uav03, uav04}. In practice, the coexistence of AirComp and communication networks is foreseeable, and the mutual interference between them is inevitable, which is particularly critical for UAV aerial base stations with limited energy and computing capabilities. In this regard, both the transmission strategy and the UAV trajectory design need to jointly consider the interference between communication and AirComp, aiming to strike a balance between communication quality and AirComp performance.
\vspace{-2mm}

\section{System Model and Problem Formulation} \label{sec:sys}
As shown in Fig.~\ref{fig:sys}, we consider a UAV-assisted integrated communication and AirComp network consisting of $M$ ground users denoted as $\mathcal{M} = \left\{1,2, \cdots, M\right\}$ and $J$ sensors denoted as $\mathcal{J} = \left\{1,2, \cdots, J\right\}$, designated for communication and AirComp, respectively. The location of user $m\in\mathcal{M}$ and sensor $j\in\mathcal{J}$ are denoted by $\bm{u}_m=[x_u^m, y_u^m, 0]$ and $\bm{v}_j=[x_v^j, y_v^j, 0]$, respectively, both located within an area denoted as $\Lambda$, as shown in Fig.~\ref{fig:sys}. These sensors are deployed to collect local data for specific missions such as sensing data (e.g., temperature, pressure, power, or humidity measurements) used for event inference. Due to limited energy storage and deployment area constraints of these distributed sensors, a UAV is dispatched as an aerial base station to fly over this region to aggregate sensing data. Meanwhile, the UAV communicates with ground users in this region to meet their uplink transmission needs, e.g. video surveillance, image transfer, and remote data processing. Considering the scarcity of spectrum resources, user data transmission and AirComp operate concurrently on the same frequency band. To reflect practical hardware constraints, we assume that the UAV and all ground devices are equipped with one antenna.

Specifically, we expatiate this process with a time-slotted model, where the UAV flight time $T$ is discretized into $N$ slots, denoted by $\mathcal{N} = \left\{1,2,\cdots,N\right\}$, each with a duration of $\delta = T/N$. In this regard, discrete waypoints serve as approximations for traversing corresponding time slots. We consider that the UAV flies at a fixed altitude $H$. The ground projections of discrete waypoints are denoted by $\bm{q}=\left\{\bm{q}_n\right\}_{n=0,1,\cdots,N}$ where $\bm{q}_n=[x^u_n, y^u_n, H]$, with $x^u_n$ and $y^u_n$ being $\textit{x}$- and $\textit{y}$-coordinates at slot $n$, respectively. In addition, the UAV initiates missions from a predefined starting point $\bm{q}_o$ and ends at a designated destination $\bm{q}_f$. The mobility constraints of the UAV are defined by
\begin{equation}
	\bm{q}_0 = \bm{q}_o,~\bm{q}_{N} = \bm{q}_f,
\end{equation}
\begin{equation}
	\left\|\bm{q}_n-\bm{q}_{n-1}\right\| \leq V_{\max}\delta,~\forall n\in\mathcal{N},
\end{equation}
where $V_{\max}$ is the maximum speed of the UAV.

In this paper, we focus on the uplink transmission problem, where sensor data aggregation and user uplink data transmission are carried out simultaneously over the same frequency band, resulting in mutual interference. In each slot, users communicate with the UAV using the time division multiple access protocol, while the UAV concurrently receives all data from both ground sensors and users.
We assume that the channels from ground users to UAV, and from sensors to UAV, are the probabilistic LoS channels~\cite{los}. Then, the channel from user $m$ to the UAV in slot $n$ is given by
\begin{equation} \label{eq:gain}
	\begin{split}
		G_{m,n} =\:&20\log_{10}\left(\frac{4\pi f_c}{c}\sqrt{\left\|\bm{q}_n-\bm{u}_{m}\right\|^2}\right)\\
		 &+ P_{\mathsf{LoS}} \mu_{\mathsf{LoS}} + P_{\mathsf{NLoS}} \mu_{\mathsf{NLoS}},~\forall m\in\mathcal{M},
	\end{split}
\end{equation}
where $f_c$ is the carrier frequency, $c$ is the light speed, $\mu_{\mathsf{LoS}}$ and $\mu_{\mathsf{NLoS}}$ are the additional losses caused by the LoS and non-LoS link, respectively, which depend on the different environments, density, and height of buildings. $P_{\mathsf{LoS}}$ is the probability of Line-of-Sight (LoS) link between user $m$ and the UAV, $P_{\mathsf{NLoS}}=1-P_{\mathsf{LoS}}$. $P_{\mathsf{LoS}}$ is given by
\begin{equation}\label{eq:plos}
P_{\mathsf{LoS}} = \frac{1}{1 + A\exp\left(-B({\frac{180}{\pi}\arcsin\frac{H}{\sqrt{\left\|\bm{q}_n-\bm{u}_{m}\right\|^2}}}-A)\right)},
\end{equation}
where $ A $ and $ B $ depend on the propagation environment. Accordingly, the channel power gain between user $m$ and the UAV in slot $n$ is given by
\begin{equation}\label{eq:gmn}
	g_{m,n} = 10^{-G_{m,n} / 10},~\forall m \in \mathcal{M},~\forall n \in \mathcal{N}.
\end{equation}
In this paper, considering that users and sensors are distributed within the same environmental area, we assume that the channel model for the sensor-to-UAV link is the same as that for the user-to-UAV link. By substituting the distance $\sqrt{\left\|\bm{q}_n-\bm{v}_{j}\right\|^2}$ between sensor $j$ and the UAV at time slot $n$ into~(\ref{eq:gain}) to~(\ref{eq:gmn}), we can derive the corresponding channel gain, which we define as $h_{j,n}$, $j \in \mathcal{J}, n \in \mathcal{N}$.

To alleviate interference among users and improve transmission efficiency and reliability, we assume that UAV communicates with only one user per timeslot. We define the user scheduling as a set of binary variables $\left\{a_{m,n}| m \in \mathcal{M}, n \in \mathcal{N}\right\}$. Then, after the user and all sensors simultaneously send their signals to the UAV over the same radio channel in the slot $n$, the received signal at the UAV is given by
\begin{equation} \label{eq:yr}
	\hat{s}_n = \sum_{j \in \mathcal{J}}{b_{j,n} h_{j,n} o_{j,n}} + \sum_{m \in \mathcal{M}}{a_{m,n}g_{m,n} \sqrt{p_{m,n}}x_{m,n}} + z_n,
\end{equation}
where $b_{j,n}\in \mathbb{C}$ is the transmit precoding coefficient controlling signal amplitude and phase at sensor $j$, $o_{j,n}\in \mathbb{C}$ is the pre-processed signal of sensor $j$, $a_{m,n}$ is the user scheduling variable of user $m$, $p_{m,n}$ is the transmit power of user $m$, $x_{m,n}$ is the transmitted signal of user $m$, and $z_n$ is the additive white Gaussian noise, i.e., $z_n\thicksim \mathcal{CN}\left(0, \sigma^2\right)$. To facilitate the power-control design and reduce transmit power, we assume that the symbols from AirComp are independent and normalized with zero mean and unit variance, i.e., $\mathbb{E}[o_{j,n}] = 0$, $\mathbb{E}[o_{j,n} o^*_{j,n}] = 1$, and so are the communication symbols for the users. Meanwhile, to focus on handling interference between users and sensors, the data among sensors are uncorrelated~\cite{uncor, gnn}, i.e., $\mathbb{E}[o_{j,n} o^*_{l,n}]=0, j \neq l$. Each device has a transmit power budget, the peak transmission power constraints for sensor $j$ and user $m$ are respectively given by
\begin{equation}
	|b_{j,n}|^2 \leq p_b^{\max},~\forall j\in\mathcal{J},~\forall n\in\mathcal{N},
\end{equation}
\begin{equation}
	0 \leq p_{m,n} \leq p^{\max},~\forall m\in\mathcal{M},~\forall n\in\mathcal{N},
\end{equation}
where $p_b^{\max}$ and $p^{\max}$ are the maximum transmission power of sensor $j$ and user $m$, respectively.

\subsection{AirComp Model}
For the proposed UAV-assisted integrated communication and AirComp network, the AirComp part in Fig.~\ref{fig:sys} shows the data aggregation process from sensors to the UAV using AirComp. Each sensor records data in each slot, such as ambient environment parameters or biological movement data. For sensor $j$, a symbol is generated by its measurement, denoted as $d_{j,n}\in \mathbb{C}$ in time slot $n$. Then, each sensor pre-processes its own signal and sends it to the UAV simultaneously. The pre-processing signal transmitted by sensor $j$ at slot $n$ can be denoted as
\begin{equation}
	o_{j,n} = \psi_{j,n} \left(d_{j,n}\right),
\end{equation}
where $\psi_{j,n}(\cdot)$ is the pre-processing function at sensor $j$ at slot $n$. For data aggregation at the UAV, the sum of the distributed signals from all sensors corresponds to the superposition that naturally occurs in the channel, and the UAV aims at computing a nomographic function of the distributed signals~\cite{nomo}. Specifically, the computation output at slot $n$, called the target function is given by
\begin{equation} \label{eq:F}
	F\left(d_{1,n}, d_{2,n}, \cdots, d_{J,n}\right) = \Xi \left(\sum_{j \in \mathcal{J}}{\psi_{j,n} \left(d_{j,n}\right)}\right),
\end{equation}
where $\Xi(\cdot)$ represents the post-processing function at the UAV. Without loss of generality, we consider the average of distributed data from $J$ sensors as the post-processing at the UAV which is expressed as $F=\frac{1}{J}\left(\sum_{j \in \mathcal{J}}{\psi_{j,n} \left(d_{j,n}\right)}\right)$. Then, upon receiving the signal, the UAV performs post-processing and scaling of the data to obtain the required estimated average, given by
\begin{equation} \label{eq:Fhat}
	\hat{F} = \frac{\eta_n \cdot \hat{s}_{n}}{J},
\end{equation}
where $\eta_n$ is a receive normalizing factor at the UAV for signal power compensation and noise suppression at slot $n$.

Due to channel fading, interference, and noise significantly affecting the computational accuracy of AirComp, we use the widely adopted MSE metric between the estimated function value $\hat{F}$ in~(\ref{eq:Fhat}) and the true value $F$ in~(\ref{eq:F}) to quantify computational errors, ensuring the proper functioning of AirComp and facilitating subsequent system optimizations. In particular, MSE at slot $n$ can be obtained by
\begin{equation} \label{eq:md}
	\begin{split}
		\text{MSE}_n &= \mathbb{E}\left[|\hat{F}- F|^2\right] \\
		&= \frac{1}{J^2}\mathbb{E}\left[\left| \eta_n \cdot \hat{s}_{n}- \sum_{j \in \mathcal{J}}{o_{j,n}}\right| ^2\right],
	\end{split}
\end{equation}
where the expectation $\mathbb{E}[\cdot]$ is taken for the randomness of the original signals $o_{j,n}$, $x_{m,n}$, and the receiver noise $z_n$. Then, substituting~(\ref{eq:yr}) into~(\ref{eq:md}), the MSE of AirComp can be further derived as
\begin{equation}
	\begin{split}
		\text{MSE}_n =~&\frac{1}{J^2}\mathbb{E}\left[\left| \sum_{j \in \mathcal{J}}{\left(\eta_n b_{j,n} h_{j,n}-1\right) o_{j,n}}\right.\right.\\
		&\left. \left. + \sum_{m \in \mathcal{M}}{\eta_n a_{m,n} g_{m,n}\sqrt{p_{m,n}}x_{m,n}} + z_n\right| ^2\right] \\
		=~&\frac{1}{J^2}\left(\sum_{j \in \mathcal{J}}\left| \eta_n b_{j,n} h_{j,n} - 1\right| ^2\right.\\
		&\left.  + \left|\eta_n\right|^2 \left(\sum_{m \in \mathcal{M}}{a_{m,n}\left| g_{m,n}\right|^2 p_{m,n}} + \sigma^2 \right)\right),
	\end{split}
\end{equation}
where $\sum_{m \in \mathcal{M}}{a_{m,n}\left| g_{m,n}\right|^2 p_{m,n}}$ is the user interference received at the UAV when it performs AirComp operations at time slot $n$, and $\sigma^2$ is the background noise power.
\subsection{Transmission Model}
For the uplink transmission from users to UAV, based on~(\ref{eq:yr}), the transmission rate of user communication during slot $n$ can be obtained as
\begin{equation}\label{eq:rate}
	R_{n} = \sum\limits_{m \in \mathcal{M}}a_{m,n}\log_2\left(1 + \frac{\left|\eta_n g_{m,n}\right|^2 p_{m,n}}{\sum\limits_{j \in \mathcal{J}}\left|{\eta_n b_{j,n} h_{j,n}}\right| ^2 + \left|\eta_n\right|^2 \sigma^2 }\right),
\end{equation}
where $\sum_{j \in \mathcal{J}}\left|{\eta_n b_{j,n} h_{j,n}}\right| ^2$ represents the interference power from AirComp and unit bandwidth is assumed without loss of generality. Considering that the UAV can serve at most one user in each slot. The constraints of scheduling strategy $a_{m, n}$, $m\in \mathcal{M}$, $n\in \mathcal{N}$ are as follows
\begin{equation}
	a_{m,n}\in \left\{0,1\right\},~\forall n \in \mathcal{N},~\forall m \in \mathcal{M},
\end{equation}
\begin{equation}
	\sum_{m=1}^{M}a_{m,n}\leq1,~\forall n \in \mathcal{N}.
\end{equation}
In addition, to ensure fairness in the user scheduling strategy, we introduce a constraint on the total number of slots available for user access to the UAV, given as
\begin{equation}
	\lfloor N/M \rfloor \leq \sum_{n \in \mathcal{N}}a_{m,n},~\forall m \in \mathcal{M},
\end{equation}
where the $\lfloor \cdot \rfloor$ is the floor function.

\subsection{Problem Formulation} \label{sec:prob}
For the considered UAV-assisted integrated communication and AirComp network, high user transmission rates, as well as low computation errors of AirComp are preferable.
However, in some practical applications, AirComp computational accuracy becomes sufficient for event inference when it reaches a certain threshold. Beyond the accuracy threshold, further improvements in computational accuracy would not alter the event inference outcomes or provide additional performance gains for the AirComp system.
Such as forest fire monitoring and smart cities, distributed data aggregation by AirComp only needs to ensure that computational errors are controlled within acceptable limits relative to actual values, without needing to achieve zero error, as long as it does not affect event inference results. In comparison, high user transmission throughput is more important, which can enhance user experience. Due to mutual interference between communication and AirComp, there exists an important trade-off between them. Meanwhile, given the dispersed distribution of users and sensors, deploying UAV-assisted communication and AirComp is necessary to provide efficient and stable transmission links.

Based on the above discussion, to meet the computational error requirements of AirComp and achieve high-quality user transmission, we jointly optimize user transmission power, sensor transmission coefficients, UAV normalizing factor, user scheduling strategy, and UAV trajectory. This optimization aims to maximize user transmission rate while satisfying the MSE threshold constraints for AirComp. Mathematically, the optimization problem is formulated as follows:
\begin{IEEEeqnarray}{cl}
	\IEEEyesnumber\label{eq:p0} \IEEEyessubnumber*
	\max_{\begin{subarray}{c}
			\big\{\left\{b_{j,n}\right\}_{j \in \mathcal{J}}, \big.\\
			\big. \left\{p_{m,n}\right\}_{m \in \mathcal{M}},\big.\\
			\big. \left\{a_{m,n}\right\}_{m \in \mathcal{M}},\big.\\
			\big. \eta_n, \bm{q}_n\big\}_{n \in \mathcal{N}}
	\end{subarray}}\,& \sum_{n \in \mathcal{N}}{R_{n}}  \\ \nonumber
	\rm{s.t.}\,& \frac{1}{J^2}\bigg(\left|\eta_n\right|^2 \left(\sum_{m \in \mathcal{M}}{a_{m,n}\left| g_{m,n}\right|^2 p_{m,n}} + \sigma^2\right)\bigg.\\
	&\bigg. + \sum_{j \in \mathcal{J}}\left| \eta_n b_{j,n} h_{j,n} - 1\right| ^2\bigg) \leq \Gamma,\forall n\in\mathcal{N}, \label{eq:mse} \\ 
	 &|b_{j,n}|^2 \leq p_b^{\max},~\forall j\in\mathcal{J},~\forall n\in\mathcal{N}, \label{eq:bjn} \\
	 &0 \leq p_{m,n} \leq p^{\max},~\forall m\in\mathcal{M},~\forall n\in\mathcal{N},\label{eq:pmn} \\
	 &0 \leq \eta_n,~\forall n\in\mathcal{N},\label{eq:eta0} \\
	 &a_{m,n}\in \left\{0,1\right\},~\forall n \in \mathcal{N},~\forall m \in \mathcal{M},\label{eq:a1} \\
	&\sum_{m=1}^{M}a_{m,n}\leq1,~\forall n \in \mathcal{N},\label{eq:a2} \\
	&\lfloor N/M \rfloor \leq \sum_{n \in \mathcal{N}}a_{m,n},~\forall m \in \mathcal{M},\label{eq:a3} \\
	&\left\|\bm{q}_n-\bm{q}_{n-1}\right\| \leq V_{\max}\delta,~\forall n \in \mathcal{N},\label{eq:traj1} \\
	&\bm{q}_0 = \bm{q}_o,~\bm{q}_{N+1} = \bm{q}_f,\label{eq:traj2}
\end{IEEEeqnarray}
where $\Gamma$ is the accuracy threshold of AirComp. The constraint in~(\ref{eq:mse}) is the threshold constraint for MSE, implying that the accuracy of AirComp must meet the predefined precision. The constraints in~(\ref{eq:bjn}) and~(\ref{eq:pmn}) impose the upper limits on transmission power for the sensors and the users, respectively.

For the formulated problem, the considered factors affect system performance in a coupled and complicated manner with three-fold difficulties. First, due to the integer scheduling variables $a_{m, n}$, $m\in \mathcal{M}$, $n\in \mathcal{N}$, the problem in~(\ref{eq:p0}) involves an integer objective function and integer constraints in~(\ref{eq:mse}),~(\ref{eq:a1}),~(\ref{eq:a2}), and~(\ref{eq:a3}). Second, due to mutual interference between communication and AirComp, the continuous variables $\left\{p_{m,n}\right\}_{m \in \mathcal{M},n \in \mathcal{N}}$, $\left\{b_{j,n}\right\}_{j \in \mathcal{J},n \in \mathcal{N}}$, and $\left\{\eta_n, \bm{q}_n\right\}_{n \in \mathcal{N}}$ are coupled with integer variables $\left\{a_{m,n}\right\}_{m \in \mathcal{M},n \in \mathcal{N}}$, leading to the non-convexity of the objective function and constraint in~(\ref{eq:mse}). Moreover, the problem involves the slotted transmission and user scheduling under a given topology, as well as the flying trajectory over the entire time horizon. Thus, the problem in~(\ref{eq:p0}) is a non-convex problem with mixed-integer nonlinear programming (MINLP), which is cumbersome to reach the optimum.

Revisiting the formulated problem, we can observe that the transmission power of users, the transmission coefficients of sensors, and the normalizing factor can be investigated in a slot-wise manner with a fixed user scheduling strategy. In contrast, user scheduling strategy and the UAV trajectory affect the system performance across all slots, and the strategies between each slot influence each other. Therefore, this problem can be decomposed into two layers where the outer layer solves the user scheduling strategy and the UAV trajectory planning, while the inner layer for the transmission strategy and the normalizing factor optimization. The inner problem can be solved independently in each slot where the transmission strategy and the normalizing factor optimization correspond to the maximization of the user transmission rate in~(\ref{eq:rate}). The outer problem is solved through reinforcement learning processing for user scheduling and the UAV trajectory.

\section{Inner Problem Solving for Transmission and Computation Strategy} \label{sec:inner}
In this section, we consider the inner problem for transmission strategy and normalizing factor optimization, with a fixed user scheduling strategy and UAV trajectory at the outer layer. The inner problem can be investigated in a slot-wise manner, aiming to meet the MSE threshold requirements for each slot while maximizing the user transmission rate. In particular, the inner problem at slot $n$ can be described as follows
\begin{IEEEeqnarray}{cl}
	\IEEEyesnumber\label{eq:p1}
	\max_{\begin{subarray}{c}
		\left\{\eta_n, \left\{b_{j,n}\right\}_{j \in \mathcal{J}}, \left\{p_{m,n}\right\}_{m \in \mathcal{M}}\right\}
	\end{subarray}} \quad & {R_{n}}  \\ 
	\rm{s.t.} \quad &(\text{\ref{eq:mse}}), (\text{\ref{eq:bjn}}), (\text{\ref{eq:pmn}}), (\text{\ref{eq:eta0}}). \label{eq:p1c}  \nonumber
\end{IEEEeqnarray}
However, the problem in~(\ref{eq:p1}) remains challenging to solve due to the non-concave objective function involving the coupled variables $p_{m,n}$, $b_{j,n}$, and $\eta_n$. Therefore, we propose an alternating optimization framework to solve the problem in~(\ref{eq:p1}), by iteratively optimizing one of the transmission power $\left\{\left\{p_{m,n}\right\}_{m \in \mathcal{M}}, \left\{b_{j,n}\right\}_{j \in \mathcal{J}}\right\}$ and the normalizing factor $\eta_n$ with the other being fixed at each iteration, detailed below.

For given the normalizing factor $\eta_n$, the problem in~(\ref{eq:p1}) can be transformed to the following problem
\begin{IEEEeqnarray}{cl}
	\IEEEyesnumber\label{eq:inn1}
	\max_{\begin{subarray}{c}
		\left\{\left\{b_{j,n}\right\}_{j \in \mathcal{J}}, \left\{p_{m,n}\right\}_{m \in \mathcal{M}}\right\}
	\end{subarray}} \quad & {R_{n}}  \label{eq:inn1o}\\ 
	\rm{s.t.} \quad &(\text{\ref{eq:mse}}), (\text{\ref{eq:bjn}}), (\text{\ref{eq:pmn}}).\label{eq:inn1c}  \nonumber
\end{IEEEeqnarray}
For given $\left\{\left\{p_{m,n}\right\}_{m \in \mathcal{M}}, \left\{b_{j,n}\right\}_{j \in \mathcal{J}}\right\}$, we optimize the normalizing factor $\eta_n$. Then, the problem in~(\ref{eq:p1}) is transformed into a feasibility check problem, which can be written as
\begin{IEEEeqnarray}{cl}
	\IEEEyesnumber\label{eq:inn2}
	\text{find} \quad & \eta_n \\
	\rm{s.t.} \quad & (\text{\ref{eq:mse}}), (\text{\ref{eq:eta0}}). \label{eq:inn2c}   \nonumber
\end{IEEEeqnarray}
We then discuss the two subproblems of the aforementioned inner-layer problem in detail in the following.

\textit{1) Transmission Strategy Optimization:}
The problem in~(\ref{eq:inn1}) remains challenging to solve because the objective function is not jointly concave with respect to $\left\{p_{m,n}\right\}_{m \in \mathcal{M}}$ and $\left\{b_{j,n}\right\}_{j \in \mathcal{J}}$, which are non-trivially coupled in the form of $p_{m,n}/{\sum_{j \in \mathcal{J}}\left|{b_{j,n}}\right| ^2}$. For the former issue, we can obtain the following inequality by introducing an auxiliary variable $\Psi_{n}$ such that,
\begin{equation}\label{eq:psi}
	\sum_{j \in \mathcal{J}}\left|{\eta_n b_{j,n} h_{j,n}}\right| ^2 + \left|\eta_n\right|^2 \sigma^2  \leq \Psi_n,~\forall n \in \mathcal{N}.
\end{equation}
Then, by replacing $\sum_{j \in \mathcal{J}}\left|{\eta_n b_{j,n} h_{j,n}}\right| ^2 + \left|\eta_n\right|^2 \sigma^2$ in~(\ref{eq:inn1o}) with the upper bound $\Psi_{n}$, we convert~(\ref{eq:inn1o}) into the following optimization problem
\begin{IEEEeqnarray}{cl}
	\IEEEyesnumber\label{eq:p_pb1}
	\max_{\begin{subarray}{c}
			\big\{\left\{b_{j,n}\right\}_{j \in \mathcal{J}}, \big.\\
			\big. \left\{p_{m,n}\right\}_{m \in \mathcal{M}}, \big.\\
			\big. \Psi_{n}\big\}
	\end{subarray}}~&{\sum\limits_{m \in \mathcal{M}}a_{m,n}\log_2\left(1 + \frac{\left|\eta_n g_{m,n}\right|^2 p_{m,n}}{\Psi_n}\right)}  \label{eq:pb_ob1} \\
	\rm{s.t.}~& \sum_{j \in \mathcal{J}}\left|{\eta_n b_{j,n} h_{j,n}}\right| ^2 + \left|\eta_n\right|^2 \sigma^2  \leq \Psi_n. \nonumber
\end{IEEEeqnarray}
However, the problem in~(\ref{eq:p_pb1}) is still difficult to solve since the objective function is not jointly concave with respect to $p_{m,n}$ and $\Psi_{n}$. Toward this issue, the objective function in~(\ref{eq:p_pb1}) can be rewritten as
\begin{equation}\label{eq:p_pb1_r}
		{R_{n}} = \frac{1}{\ln2}\sum_{m \in \mathcal{M}}a_{m,n}\left[\ln\left(\left|\eta_n g_{m,n}\right|^2 p_{m,n} + \Psi_n\right) - \ln\Psi_n\right].
\end{equation}
Furthermore, we resort to~\cite[Lemma 1]{lem1} to obtain the following equality
\begin{equation}
-\ln\left(\Psi_n\right) = \max_{t_n>0} {\left\{-t_n\Psi_n+\ln t_n+1\right\}},
\end{equation}
by introducing auxiliary variable $t_n$. Therefore, the expression in~(\ref{eq:p_pb1_r}) can be rewritten as
\begin{equation}
	\begin{split}
		R_n =~&\max_{t_n>0} \Bigg\{ \frac{1}{\ln2}\sum_{m \in \mathcal{M}}a_{m,n}\bigg[\ln\left(\left|\eta_n g_{m,n}\right|^2 p_{m,n} + \Psi_n\right) \bigg.\Bigg. \\
		&\Bigg.\bigg. - t_n\Psi_n + \ln t_n + 1\bigg]\Bigg\}, \\
	\end{split}
\end{equation}
and the optimal solution is $t_n=1/\Psi_n$. Then, the problem in~(\ref{eq:inn1}) can be rewritten as
\begin{IEEEeqnarray}{cl}
	\IEEEyesnumber\label{eq:p_pb2} \IEEEyessubnumber*  \nonumber
	\max_{\begin{subarray}{c}
		\big\{\left\{b_{j,n}\right\}_{j \in \mathcal{J}}, \big.\\
		\big. \left\{p_{m,n}\right\}_{m \in \mathcal{M}}, \big.\\
		\big. \Psi_{n}, t_n\big\}	
	\end{subarray}}~& \sum_{m \in \mathcal{M}}a_{m,n}\bigg[\ln\left(\left|\eta_n g_{m,n}\right|^2 p_{m,n} + \Psi_n\right) \bigg. \\
	&\bigg. - t_n\Psi_n + \ln t_n + 1\bigg] \\
	\rm{s.t.}~& t_n>0,  \label{eq:t_con}\\
	& (\text{\ref{eq:mse}}), (\text{\ref{eq:bjn}}), (\text{\ref{eq:pmn}}), (\text{\ref{eq:psi}}). \nonumber
\end{IEEEeqnarray}
It can be shown that~(\ref{eq:p_pb2}) is convex w.r.t. either $\left\lbrace \left\{p_{m,n}\right\}_{m \in \mathcal{M}}, \left\{b_{j,n}\right\}_{j \in \mathcal{J}}, \Psi_n\right\rbrace $ or $t_n$. Consequently, it can be solved by alternating optimization. With fixed $\left\lbrace \left\{p_{m,n}\right\}_{m \in \mathcal{M}}, \left\{b_{j,n}\right\}_{j \in \mathcal{J}}, \Psi_n\right\rbrace $, the optimal $t_n$ can be derived in closed forms as
\begin{equation}\label{eq:t}
	t_n^{\ast} = 1 / \Psi_n.
\end{equation}
On the other hand, for given $t^*_n$, the optimal $\left\lbrace \left\{p_{m,n}\right\}_{m \in \mathcal{M}}, \left\{b_{j,n}\right\}_{j \in \mathcal{J}}, \Psi_n\right\rbrace $ can be obtained by solving
\begin{IEEEeqnarray}{cl}
	\IEEEyesnumber\label{eq:p_pbp}  \nonumber
	\max_{\begin{subarray}{c}
		\big\{\left\{b_{j,n}\right\}_{j \in \mathcal{J}}, \big.\\
		\big. \left\{p_{m,n}\right\}_{m \in \mathcal{M}}, \big.\\
		\big. \Psi_{n}\big\}	
\end{subarray}}~& \sum_{m \in \mathcal{M}}a_{m,n}\bigg[\ln\left(\left|\eta_n g_{m,n}\right|^2 p_{m,n} + \Psi_n\right) \bigg. \\
&\bigg. - t_n\Psi_n + \ln t_n + 1\bigg] \\
	\rm{s.t.} \quad &(\text{\ref{eq:mse}}), (\text{\ref{eq:bjn}}), (\text{\ref{eq:pmn}}), (\text{\ref{eq:psi}}). \nonumber
\end{IEEEeqnarray}
The objective in~(\ref{eq:p_pbp}) is concave with respect to $\left\{\left\{p_{m,n}\right\}_{m \in \mathcal{M}}, \left\{b_{j,n}\right\}_{j \in \mathcal{J}}, \Psi_n\right\}$, and all constraints are either linear or convex, satisfying the conditions for convex optimization. Therefore, it can be efficiently solved by using a convex optimization solver, e.g., CVX. In the above, an approximate solution to~(\ref{eq:inn1}) is obtained by alternately updating $\left\{\left\{p_{m,n}\right\}_{m \in \mathcal{M}}, \left\{b_{j,n}\right\}_{j \in \mathcal{J}}, \Psi_n\right\}$ and $t_n$. Specifically, given $\left\{\left\{p_{m,n}\right\}_{m \in \mathcal{M}}, \left\{b_{j,n}\right\}_{j \in \mathcal{J}}, \Psi_n\right\}$, we find the optimal $t^*_n$ according to~(\ref{eq:t}). Then, with the optimal $t^*_n$, find the optimal $\left\{\left\{p^*_{m,n}\right\}_{m \in \mathcal{M}}, \left\{b^*_{j,n}\right\}_{j \in \mathcal{J}}, \Psi^*_n\right\}$ by solving~(\ref{eq:p_pbp}). Alternately update $\left\{\left\{p_{m,n}\right\}_{m \in \mathcal{M}}, \left\{b_{j,n}\right\}_{j \in \mathcal{J}}, \Psi_n\right\}$ and $t_n$ until the convergence condition is satisfied.

\textit{2) Normalizing Factor Optimization:}
When the transmit power $\left(p_{m,n}, b_{j,n}\right)$ is fixed, the normalizing factor optimization problem is a feasibility check problem as~(\ref{eq:inn2}). The feasibility check problem simply requires finding a solution that satisfies all constraints. Specifically, the problem in~(\ref{eq:inn2}) is primarily affected by the quadratic constraint in~(\ref{eq:mse}). It involves identifying a value of $\eta_n$ within its domain that ensures the AirComp computational accuracy, measured by the MSE, is less than the threshold $\Gamma$. It is evident that the value of $\eta_n$ that minimizes the MSE is the solution that best satisfies the constraint in~(\ref{eq:mse}), the problem in~(\ref{eq:inn2}) can be transformed into a convex quadratic problem as follows:
\begin{algorithm}
	\caption{Alternating optimization for solving~(\ref{eq:p1})}
	\label{alg:1}
	\SetKwData{In}{\textbf{in}}\SetKwData{To}{to}
	\DontPrintSemicolon
	\SetAlgoLined
	\KwIn {$p^{\max}, b^{\max}, \Gamma, \left\{a_{m,n}\right\}_{m \in \mathcal{M}}, \bm{q}_n$.}
	\KwOut {$\left\{p_{m,n}\right\}_{m \in \mathcal{M}}, \left\{b_{j,n}\right\}_{j \in \mathcal{J}}$, $\eta_n$.}
	{Initialization:~$l \leftarrow 0$, set $\eta^{(0)}_n$}.\\
	\Repeat{$\left|\left[p^{(l)}_{m,n}-p^{(l-1)}_{m,n}\right]_{m \in \mathcal{M}}, \left[b^{(l)}_{j,n}-b^{(l-1)}_{j,n}\right]_{j \in \mathcal{J}}, \eta^{(l)}_n-\eta^{(l-1)}_n\right|$\newline
		$\leq \xi$.}{
		{Initialization:~$i \leftarrow 0$, set $\left\{\left\{p^{(0)}_{m,n}\right\}_{m \in \mathcal{M}}, \left\{b^{(0)}_{j,n}\right\}_{j \in \mathcal{J}}, \Psi^{(0)}_n\right\}$ according to~(\text{\ref{eq:mse}}), (\text{\ref{eq:bjn}}), (\text{\ref{eq:pmn}}), and~(\ref{eq:psi})}.\\
		$l \leftarrow l+1$;\\
		\Repeat{$\left|\left[p^{(i)}_{m,n}-p^{(i-1)}_{m,n}\right]_{m \in \mathcal{M}}, \left[b^{(i)}_{j,n}-b^{(i-1)}_{j,n}\right]_{j \in \mathcal{J}}\right|\leq \xi$.}{
			$i \leftarrow i+1$;\\
			Given $\left\lbrace \left\{p^{(i-1)}_{m,n}\right\}_{m \in \mathcal{M}}, \left\{b^{(i-1)}_{j,n}\right\}_{j \in \mathcal{J}}, \Psi^{(i-1)}_n, \eta^{(l-1)}_n\right\rbrace $, find the optimal $t^{i}_n$ according to~(\ref{eq:t}); \\
			Given $t^{i}_n$ and $\eta^{(l-1)}_n$, find the optimal $\left\lbrace \left\{p^{(i)}_{m,n}\right\}_{m \in \mathcal{M}}, \left\{b^{(i)}_{j,n}\right\}_{j \in \mathcal{J}}, \Psi^{(i)}_n\right\rbrace $ by solving~(\ref{eq:p_pbp}).
		}
		Given $\left\lbrace \left\{p^{(l)}_{m,n}\right\}_{m \in \mathcal{M}}, \left\{b^{(l)}_{j,n}\right\}_{j \in \mathcal{J}}, \Psi^{(l)}_n\right\rbrace $ as $\left\lbrace \left\{p^{(i)}_{m,n}\right\}_{m \in \mathcal{M}}, \left\{b^{(i)}_{j,n}\right\}_{j \in \mathcal{J}}, \Psi^{(i)}_n\right\rbrace $, obtain the solution for $\eta^{(l)}_n$ according to~(\ref{eq:eta_}).
	}
\end{algorithm}
\begin{IEEEeqnarray}{cl}
	\begin{split}
	\IEEEyesnumber\label{eq:opteta}
	\min_{\eta_n>0} \quad &\frac{1}{J^2}\left(\left|\eta_n\right|^2 \left(\sum_{m \in \mathcal{M}}{a_{m,n}\left| g_{m,n}\right|^2 p_{m,n}} + \sigma^2\right)\right. \\
	&\left. \sum_{j \in \mathcal{J}}\left| \eta_n b_{j,n} h_{j,n} - 1\right| ^2\right).
	\end{split}
\end{IEEEeqnarray}
Then, by setting the first derivative of the MSE in~(\ref{eq:mse}) to be zero, we can obtain a feasible solution
\begin{equation}\label{eq:eta_}
	\eta^*_n = \frac{\sum_{j}\left|b_{j,n} h_{j,n}\right|}{\sum_{j}\left|b_{j,n} h_{j,n}\right|^2 + \sum_{m}{a_{m,n}\left| g_{m,n}\right|^2 p_{m,n}} + \sigma^2}.
\end{equation}

Therefore, the original optimization problem in~(\ref{eq:p1}) can be tackled through an alternating optimization scheme involving~(\ref{eq:p_pb2}) and~(\ref{eq:eta_}), which is summarized in Alg.~\ref{alg:1}. Specifically, we begin by considering a fixed $\eta_n$ and obtain an approximate solution to~(\ref{eq:p_pb2}) by alternately updating optimization $\left\{p_{m,n}, b_{j,n}\right\}$ and $t_n$. Subsequently, Using the optimized $\left\{p_{m,n}^*, b_{j,n}^*\right\}$, as the fixed variables, we can obtain the optimized normalizing factor $\eta^*_n$ in~(\ref{eq:eta_}). Finally, by alternating iterations until convergence, the solution to the original problem can be obtained.

\section{Outer Learning for Trajectory and User Scheduling} \label{sec:outer}
In this section, we consider the outer problem of UAV trajectory and user scheduling. Unlike the inner problem, which can be solved at each slot, the UAV trajectory and user scheduling impact user transmission and AirComp throughout the entire mission time. The mutual influence of UAV trajectory, user scheduling strategy, and inner strategy in optimization problems leads to significant complexity in problem-solving. This complexity renders traditional optimization algorithms difficult in the solution-seeking process.

In this regard, DRL emerges as an attractive solution~\cite{sgdrl}. However, for our considered problem with high-dimensional state space and continuous action space, the traditional DRL faces the following problems~\cite{dedrl}. On one hand, on-policy algorithms like proximal policy optimization (PPO) utilize the same policy for both generating samples and updating network parameters, necessitating new samples for each gradient step, which leads to low sampling efficiency. On the other hand, off-policy algorithms such as deep deterministic policy gradient (DDPG) enhance sample utilization by reusing past experiences. However, they are highly sensitive to hyperparameter settings, including learning rates and exploration constants, requiring careful tuning for optimal results and maintaining robustness. Additionally, off-policy methods are not inherently compatible with traditional policy gradient formulations. Based on the above discussion, we propose an SAC-based DRL algorithm to address our problem. 
SAC is a robust and stable DRL algorithm adept at solving complex problems involving continuous state and hybrid action spaces. Its stochastic policy formulation allows the discrete and continuous action components to be modeled separately yet optimized jointly, making it well-suited for optimizing the UAV trajectory and user scheduling problems considered. Compared with other DRL algorithms, SAC encourages broader exploration during policy training to capture multiple near-optimal paths, thereby enhancing the learning speed for complex tasks.
\vspace{-4.5mm}
\begin{figure*}[t]
        \centering
        \includegraphics[width=13.5cm]{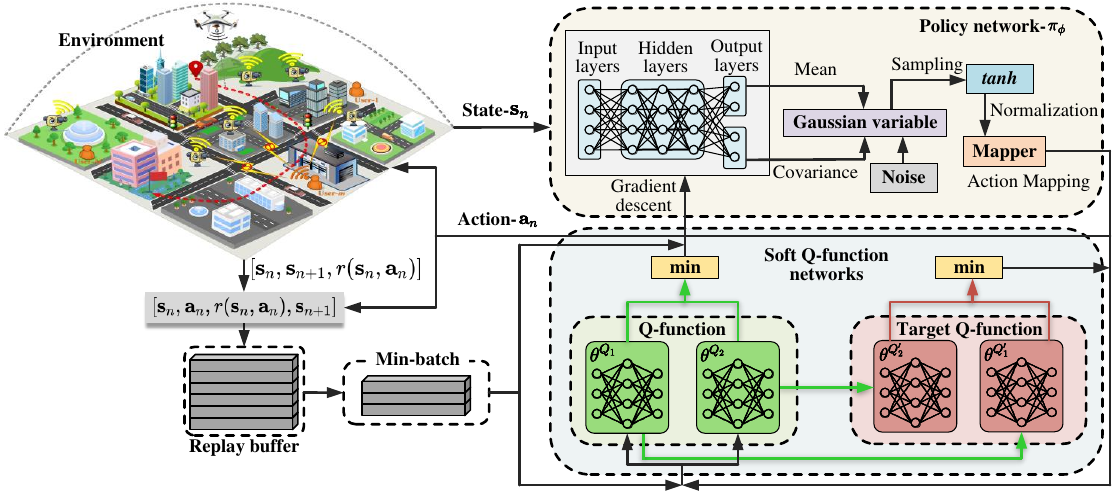}
        \caption{The framework of SAC-based DRL algorithm.}
        \vspace{-1mm}
        \label{fig:sac-tso}
\end{figure*}

\subsection{Trajectory and Scheduling as MDP} \label{sec:TS}
The UAV trajectory planning and user scheduling strategy have long-term implications on transmission and AirComp during missions, particularly in UAV trajectory planning where strategies for time slots interact. Considering fairness among users, scheduling also involves mutual interactions of strategies for time slots. To address these long-term effects, the UAV can be treated as an agent, trajectory planning and user scheduling strategy as action space, and transmission and AirComp states as state space, modeled as a Markov decision process (MDP) to track the long-term effects. Specifically from a learning perspective, the considered problem can be defined as policy search in the MDP where UAV acts as an agent, defined by a tuple $\left(\mathcal{S}, \mathcal{A}, r\right)$, where $\mathcal{S}$ is the state space, $\mathcal{A}$ is the action space, and $r$ is the reward function. At each slot $n$, with a given state $\mathbf{s}_{n} \in \mathcal{S}$, the agent selects action $\mathbf{a}_{n} \in \mathcal{A}$ with respect to its policy $\pi(\mathbf{a}_{n}|\mathbf{s}_{n})$. Then, the agent interacts with the environment by executing the action to achieve a new state $\mathbf{s}_{n+1} \in \mathcal{S}$, and the environment emits a reward $r: \mathcal{S} \times \mathcal{A}$. To be specific, the corresponding elements are defined as follows.

\textit{1) State Space $\mathcal{S}$}: The state $\mathbf{s}_{n} \in \mathcal{S}$, $n\in\mathcal{N}$, corresponds to the system environment state at slot $n$, denoted as
\begin{equation}
	\mathbf{s}_{n} = \left\{ \left[\textstyle\sum_{i=1}^{n}a_{m,i}\right]_{m\in\mathcal{M}}; R_n; \text{MSE}_n; x^u_n, y^u_n  \right\},
\end{equation}
which has a cardinality equal to $M + 4$. In state $\mathbf{s}_{n}$, $\left[\textstyle\sum_{i=1}^{n}a_{m,i}\right]_{m\in\mathcal{M}}$ is the historical scheduling frequency of each user in the slot $n$, $R_n$ is the user transmission rate at slot $n$, $\text{MSE}_n$ is the accuracy of AirComp at slot $n$, and $\left[x^u_n, y^u_n\right]$ represents the 2D position of the UAV.

\textit{2) Action Space $ \mathcal{A} $}: The action $\mathbf{a}_{n} \in \mathcal{A}$, $n\in\mathcal{N}$ is defined by a $(M+2)$-dimensional vector $\mathbf{a}_{n}=\left\{\Delta x_n, \Delta y_n;\left[a_{m,n}\right]_{m\in\mathcal{M},n\in\mathcal{N}}\right\}$, where $\left[\Delta x_n, \Delta y_n\right]$ is the change in location between two successive slots, i.e., $ \left[\Delta x_n, \Delta y_n\right] = \left[x_n-x_{n-1}, y_n-y_{n-1}\right]$.

\textit{3) Reward $r(\mathbf{s}_{n}, \mathbf{a}_{n})$}: The reward measures the effect of the action taken by an agent for a given state, which further guides the agent to find the best trajectory and scheduling strategy. To meet various requirements such as user transmission rate, fairness, AirComp accuracy, and the specified UAV destination, the reward function comprises two components: one concerns the scheduling and the other is for trajectory, and they are specified as follows.
\begin{itemize}
	\item {\em Scheduling Reward}: We aim to maximize the transmission rate of users while guaranteeing fairness through reasonable scheduling of users. On one hand, we incorporate the transmission rate of the users accessing the current slot into the reward, to find the proper scheduling solution. On the other hand, we combine the total number of time slots transmitted by the users into reward, to encourage the average transmission opportunities for all users and achieve fairness, corresponding to the constraint in~(\ref{eq:a3}). Then, at slot $n$, the scheduling reward is given by
	\begin{equation}
		r^c_{n}= \left\{\begin{array}{rcl}
			R_n, & \text{if}~(\text{\ref{eq:bjn}})-(\text{\ref{eq:a3}}),\\
			0, & \text{otherwise}. \end{array}\right.
	\end{equation}
	\item {\em Trajectory Reward}: We need to set an appropriate waypoint reward to enable the UAV to efficiently achieve transmission and AirComp while ensuring it flies from the designated starting point to the specified destination. Otherwise, the UAV may cannot reach the destination. Therefore, we design a distance and slot-based penalty. When the slot $n$ is relatively small, there is sufficient trajectory adjustment space in subsequent slots. Therefore, a smaller penalty is assigned to encourage the UAV to explore potentially better strategies. Conversely, as it approaches time slot $N$, reaching the destination promptly to satisfy constraint in~(\ref{eq:traj2}) becomes a more critical task. Therefore, we define the trajectory reward $r^d_n$ as
	\begin{equation}
		r^d_{n}= \left\{\begin{array}{rcl}
			&R_f,~\text{if}~n=N~\text{and}~\sqrt{\left\|\bm{q}_n-\bm{q}_F\right\|^2} \leq V_{\max}\delta,\\
			&-n \times \sqrt{\left\|\bm{q}_n-\bm{q}_F\right\|^2},~\text{otherwise}, \end{array}\right.
	\end{equation}
	where $R_f$ represents the arrival reward that encourages the UAV to reach the destination by the final time slot. Moreover, we explicitly address the issue of flying outside the designated area by implementing a boundary projection mechanism.
\end{itemize}

Combining communication reward $r^c_n$ and arrival reward $r^d_n$, we define the overall reward to meet multiple demands in our problem as
\begin{equation}
	r(\mathbf{s}_{n}, \mathbf{a}_{n})= \lambda_1r^c_n+\lambda_2r^d_n,
\end{equation}
where $\lambda_1$ and $\lambda_2$ are both weighted coefficients.

\subsection{SAC-based DRL Algorithm} \label{sec:SAC-TSO}
The proposed SAC-based DRL algorithm framework, as shown in Fig.~\ref{fig:sac-tso}, consists of a policy network, two soft Q-function networks, two target soft Q-function networks, and a replay buffer. Unlike traditional reinforcement learning algorithms, the SAC algorithm incorporates the concept of maximum policy entropy in the objective function, and the maximum entropy objective is as follows:
\begin{algorithm}
	\caption{SAC-Based DRL Algorithm}
	\label{alg:2}
	\SetKwData{In}{\textbf{in}}\SetKwData{To}{to}
	\DontPrintSemicolon
	\SetAlgoLined
	\KwIn {$p^{\max}, b^{\max}, \Gamma$, $\bm{q}_0$, and $\bm{q}_f$.}
	\KwOut {$\phi, \theta_{1}, \theta_{2}, \left\{p_{m,n}, b_{j,n}, \eta_{n}\right\}_{m \in \mathcal{M}, j \in \mathcal{J}, n \in \mathcal{N}}$.}
	{Initialized parameters of policy network, soft Q-function networks, and target soft Q-function networks: $\phi$, $\theta_{1}$, and $\theta_{2}$, $\theta'_{1} \gets \theta_{1}$, $\theta'_{2} \gets \theta_{2}$.}\\
	\For{\rm{each episode}}{
		Observe initial state $\mathbf{s}_1$; \;
		\For{\rm{slot} n=1,2,...,N}{
				Sample action from policy $\mathbf{a}_n \thicksim\pi_{\phi}(\mathbf{a}_n|\mathbf{s}_n)$; \;
				{\textbf{Inner Problem Solving}: }	\\
				Execute action $\mathbf{a}_n$:  For given $\left\{a_{m,n}\right\}_{m \in \mathcal{M}}$ and $\bm{q}_n$, solve~(19) via Alg.~\ref{alg:1} to obtain $\left\{p_{m,n}\right\}_{m \in \mathcal{M}}, \left\{b_{j,n}\right\}_{j \in \mathcal{J}}$ and $\eta_n$; \;
				Compute reward $r(\mathbf{s}_n, \mathbf{a}_n)$, observe next state $\mathbf{s}_{n+1} $; \;
				\If{n = N \rm{and} $\sqrt{\left\|\bm{q}_n-\bm{q}_F\right\|^2} \leq V_{\max}\delta$}{
					Achieve reward $R_f$; \;
				}
				Store $(\mathbf{s}_{n}, \mathbf{a}_{n}, r(\mathbf{s}_n, \mathbf{a}_n),\mathbf{s}_{n+1})$ into memory $\mathcal{D}$;
			}
			{\textbf{Outer Learning}: }	\\
		\If{$\mathcal{D}$ \rm{is filled with samples}}{
			Randomly select a transition batch from memory; \;
			Update soft Q-function parameters with~(\ref{eq:Q}); \;
			Update policy network parameter with~(\ref{eq:pi});\;
			Update target soft Q-function parameters with~(\ref{eq:theta1}).\;
	}}
\end{algorithm}
\begin{equation}
	\max_{\pi}\sum_{n=1}^N~\mathbb{E}_{(\mathbf{s}_{n}, \mathbf{a}_{n})\sim\rho_{\pi}}\left[\gamma^{n-1}r(\mathbf{s}_{n}, \mathbf{a}_{n})-\beta\log\left(\pi(\mathbf{a}_{n}|\mathbf{s}_{n})\right)\right],
\end{equation}
where $\rho_{\pi}$ is the state-action marginal distribution following $\pi$, $\gamma\in(0,1)$ is a discount factor, and $\beta$ is the temperature parameter that determines the weight of the entropy versus the reward. The first term $\gamma^{n-1}r(\mathbf{s}_{n}, \mathbf{a}_{n})$ is for the long-term return and the second term $-\beta\log\left(\pi(\mathbf{a}_{n}|\mathbf{s}_{n})\right)$ is the entropy of the policy distribution. This design considers both the immediate reward of the policy and its randomness, enhancing the stability and robustness of the algorithm, and achieving a balance between exploration and exploitation~\cite{sac}. Due to the incorporation of the entropy, the state-value function in SAC is referred to as the soft state-value function given as
\begin{equation}
	\mathbf{v}_{\pi}(\mathbf{s}_{n}) = \mathbb{E}_{\mathbf{a}_{n}\sim{\pi}}\left[Q_{\pi}(\mathbf{s}_{n}, \mathbf{a}_{n})-\beta\log\left(\pi(\mathbf{a}_{n}|\mathbf{s}_{n})\right)\right],
\end{equation}
where $Q_{\pi}(\mathbf{s}_{n}, \mathbf{a}_{n})$ is the state-action value (i.e., Q-value). Additionally, the introduction of double Q-function networks aims to mitigate the overestimation bias in Q-value estimation, improving the stability and accuracy of the training~\cite{dq}. For the policy design, the policy network can be regarded as a probability distribution function for actions, such as a Gaussian function, where the mean and covariance are output by a neural network. Then, the raw action is then obtained by sampling from the Gaussian random variable. Next, the raw actions need to be normalized to constrain them within a reasonable range, and then finally map to the actual action range to obtain the final actions.
\begin{table}[h]\centering
	\caption{Simulation Parameters}
	\label{table}
	\renewcommand{\arraystretch}{1.5}
	\begin{tabular}{c|c|c}
		\hline
		\textbf{Parameter} & \textbf{Description} & \textbf{Value} \\
		\hline
		$H$ & Altitude of UAV  & 100~m \\
		$V_{\max}$ & Maximum speed of UAV  & 30~m/s \\
		$p^{\max}$  & Maximum power of each user & 0.2~W \\
		$b^{\max}$  & Maximum power of each sensor & 0.05~W \\
		$T$  & Mission duration & 60~s \\
		$\delta$  & Duration of each slot & 1~s \\
		$\Gamma$  & MSE threshold & 0.015 \\
		$\sigma^2$ & Noise power & -95~dBm \\
		$f_c$ & Carrier frequency & 2~GHz \\
		$(A, B)$ & Environment factor  & (9.613, 0.158) \\
		$(\mu_{\mathsf{LoS}}, \mu_{\mathsf{NLoS}})$ & LoS and NLoS attenuation  & (1~dB, 20~dB) \\
		$\xi$ & Convergence accuracy of iterations  & 0.001  \\  
		$\alpha_{q}$ & Soft Q-function learning rate  & 1$\times$10{\textsuperscript{-4}}  \\
		$\alpha_{\pi}$ & Policy learning rate  & 1$\times$10{\textsuperscript{-4}}  \\ 
		$\gamma$ & Discounted factor  & 0.90  \\
		min-batch & Batch size  & 64  \\     
		$\left| \mathcal{D} \right|$ & Buffer capacity  & 10{\textsuperscript{6}}  \\
		\hline        
	\end{tabular} 
\end{table}

The proposed algorithm design is detailed in Alg.~\ref{alg:2}. The critical procedures are explained as follows.
First, initialize all neural networks with parameters $\left\{\theta_1, \theta_2, \theta'_1, \theta'_2, \phi\right\} $ for the two soft Q-function networks, two target soft Q-function networks, and a policy network, respectively. Also, initialize the environment parameters and state before the start of each episode. Secondly, the agent selects and executes action $\mathbf{a}_{n}$ according to the state $\mathbf{s}_{n}$, and obtains the immediate reward $r(\mathbf{s}_n, \mathbf{a}_n)$ with an updated observation $\mathbf{s}_{n+1}$. Then, $(\mathbf{s}_{n}, \mathbf{a}_{n}, r(\mathbf{s}_n, \mathbf{a}_n),\mathbf{s}_{n+1})$ is stored as a transition in the reply buffer $\mathcal{D}$. After sufficient training, $D$ groups of transitions $(\mathbf{s}_{n}^{d}, \mathbf{a}_{n}^{d}, r^{d}(\mathbf{s}_n, \mathbf{a}_n), \mathbf{s}_{n+1}^{d})$ are randomly selected from $\mathcal{D}$ for learning. The parameters of soft Q-function network $\theta_i,i=1,2,$ are updated by minimizing the soft Bellman residual defined as
\begin{equation} \label{eq:Q}
\begin{aligned}
	&J_{Q}\left(\theta_{i}\right) =\mathbb{E}_{\left(\mathbf{s}_{n}, \mathbf{a}_{n}\right) \sim \mathcal{D}}\Bigg[\frac { 1 } { 2 } \bigg(Q_{\theta_{i}}\left(\mathbf{s}_{n}, \mathbf{a}_{n}\right)-\bigg(r\left(\mathbf{s}_{n}, \mathbf{a}_{n}\right)\bigg.\bigg.\Bigg. \\
	& \Bigg.\bigg.\bigg.+\gamma\left(\min_{i=1,2} Q_{\theta'_{j}}\left(\mathbf{s}_{n+1}, \mathbf{a}_{n+1}\right)-\beta \log \pi_{\phi}\left(\mathbf{a}_{n+1} \mid \mathbf{s}_{n+1}\right)\right)\bigg)\bigg)^{2}\Bigg].
\end{aligned}
\end{equation}
The parameter of policy function $\phi$ is updated by
\begin{equation} \label{eq:pi}
\begin{aligned}
	J_{\pi}(\phi)=~&\mathbb{E}_{\mathbf{s}_{n} \sim \mathcal{D}, \epsilon_{n}}\Big[\beta \log \pi_{\phi}\left(f_{\phi}\left(\epsilon_{n} ; \mathbf{s}_{n}\right) \mid \mathbf{s}_{n}\right)\Big. \\
	& \Big.-\min_{i=1,2} Q_{\theta_{i}}\left(\mathbf{s}_{n}, f_{\phi}\left(\epsilon_{n} ; \mathbf{s}_{n}\right)\right)\Big] .
\end{aligned}
\end{equation}
In~(\ref{eq:pi}), the reparameterization trick is employed as the solution for policy gradient, in which the policy is rewritten as $\mathbf{a}_n = f_{\phi}\left(\epsilon_{n} ; \mathbf{s}_{n}\right)$, where $\epsilon_{n}$ is an independent noise vector. We can solve the optimal $\beta$ after solving for $Q^*_{\theta_{i}}$ and $\pi^*_\phi$:
\begin{equation} \label{eq:beta}
\beta^{*}=\arg \min _{\beta} \mathbb{E}_{\mathbf{a}_{n} \sim \pi_\phi^{*}}\left[-\beta \log \pi_{\phi}^{*}\left(\mathbf{a}_{n} \mid \mathbf{s}_{n} ; \beta\right)-\beta \overline{\mathcal{H}}\right],
\end{equation}
where $\overline{\mathcal{H}}=-\prod\limits_{\overline{m}=1}^{M+2}\left(\varpi_{\overline{m}}\right)$ is the target entropy based on the action space dimension, and $\varpi_{\overline{m}}$ is the space dimension of the $\overline{m}$-th action. Finally, the parameters of the target soft Q-function network are updated as
\begin{equation} \label{eq:theta1}
  \theta'_{i} \gets \tau \theta_{i} + \left(1 - \tau\right)\theta'_{i}, \quad \tau\ll 1,
\end{equation}
according to the soft update rule, which helps improve the stability of learning.
\vspace{-2mm}

\subsection{Complexity Analysis} \label{sec:Compl}
\textit{1) Complexity Analysis of Alternating Optimization}: Each iteration in the alternating optimization procedure consists of two main steps, corresponding to the optimization problem in~(\ref{eq:p_pb2}) and a closed-form solution in~(\ref{eq:eta_}). Disregarding the closed-form solution in the iterative process, we first analyze the primary computation complexity incurred by the problem in~(\ref{eq:p_pb2}). Generally, solving a convex optimization with the interior‐point method corresponds to the complexity of $\mathcal{O}\left(n^{3.5}\log\left(\tfrac1\epsilon\right)\right)$,
where $n$ is the number of optimization variables and $\epsilon$ is the required precision. Then, the problem in~(\ref{eq:p_pbp}) incorporates $J+M$ decision variables and 1 slack variables, and thus the magnitude of complexity is $\mathcal{O}\left(\left(J+M\right)^{3.5}\log\left(\tfrac1{\epsilon_1}\right)\right)$, with $\epsilon_1$ being the computation precision. Further, assume $L_1$ and $L_2$ iterations are required to solve the transmission strategy optimization and normalizing factor problem, respectively. Then, the computation complexity of the inner layer algorithm is given as $\mathcal{O}\left(L_1L_2\left(\left(J+M\right)^{3.5}\log\left(\tfrac1{\epsilon_1}\right)\right)\right).$

\vspace{1.5mm}
\textit{2) Complexity Analysis of SAC-Based algorithm}: The complexity of the SAC-based algorithm can be divided into training and execution complexity, and both of them depend on the number of neurons and layers in the deep neural network. SAC involves the policy network, as well as the training and target soft Q-function networks. Among them, the policy network contains an input layer, an output layer, and two hidden layers with $L_i^{p}$ neurons in the $i$-th layer. The number of neurons in the input layer and output layer of the policy network is $L_0^{p} = M+4$, $L_4^{p} = M+2$, respectively. Similarly, the number of neurons in the input layer and output layer of the soft Q-function network is $L_0^{Q} = 2M+6$, $L_4^{Q} = 1$, respectively.

In the training process, the policy network and soft Q-function networks need to be trained. Besides, the training process needs the prediction results from the target soft Q-function networks. Thus, a single back-propagation training step for the proposed SAC structure will contribute to the complexity, given as $\mathcal{O}\left(\sum_{0}^{2}L_i^{p}L_{i+1}^{p}+4\sum_{0}^{2}L_i^{Q}L_{i+1}^{Q}\right).$ In addition, until the replay memory $\mathcal{D}$ is filled with samples, the agent is only explored and not trained. Therefore, the complexity in training can be calculated as
\begin{equation}
	\begin{aligned}
		\mathcal{O}\Bigg(&\left(N\Xi-|\mathcal{D}|\right)D\left(\textstyle\sum_{0}^{2}L_i^{p}L_{i+1}^{p}+4\textstyle\sum_{0}^{2}L_i^{Q}L_{i+1}^{Q}\right) \\
		&+ N\Xi\left(\textstyle\sum_{0}^{2}L_i^{p}L_{i+1}^{p}+L_1L_2\left(\left(J+M\right)^{3.5}\log\left(\tfrac{1}{\epsilon_1}\right)\right)\right)\Bigg),
	\end{aligned}
\end{equation}
where $\Xi$ and $D$ are the total number of training episodes and min-bach, respectively. During online execution, the UAV performs only the policy network and the alternating optimization in each time slot (step). Therefore, the execution complexity is given as $\mathcal{O}\left(N\left(\sum_{0}^{2}L_i^{p}L_{i+1}^{p}+L_1L_2\left(\left(J+M\right)^{3.5}\log\left(\tfrac1{\epsilon_1}\right)\right)\right)\right).$

\vspace{-0.5mm}
\section{Simulation Results} \label{sec:sim}
In this section, we present the simulation results to show the performance. We consider an area of 1,000~m $ \times $ 1,000~m. There are 15 users, 36 sensors, and a UAV. The devices are randomly located within the area. The policy network and soft Q-function networks each have two fully connected hidden layers, with 128 neurons in each layer. One episode consists of $N$ epochs, and 4000 episodes are used to train the networks. The main simulation parameters are summarized in Table~\ref{table}, used as defaults unless otherwise noted.
\vspace{-2.5mm}
\subsection{Convergence of SAC-based DRL algorithm}
\begin{figure}[t]
        \centering
        \includegraphics[width=8cm]{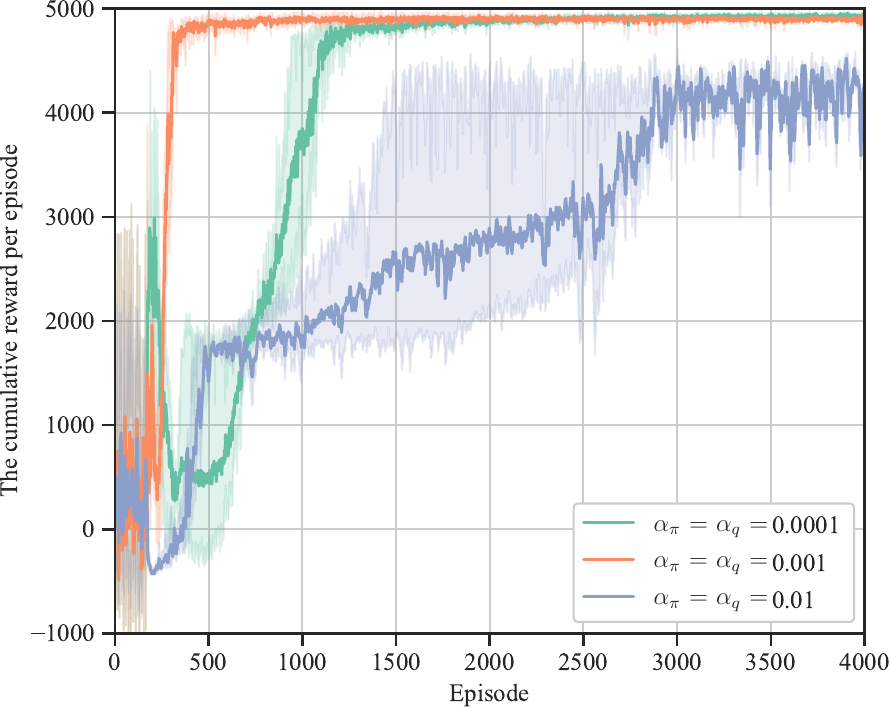}
        \vspace{-1.5mm}
        \caption{Cumulative reward with different learning rate of the neural networks.}
        \vspace{-4.5mm}
        \label{fig:conver_r}
\end{figure}

We first demonstrate the convergence of the proposed SAC-based approach in Fig.~\ref{fig:conver_r}. It shows the cumulative reward convergence of the SAC-based algorithm under different learning rates of the policy network and the soft Q-function network. It is evident that the training process eventually converges. The convergence trend of the curves and the variance indicated by the shaded areas reveal that the learning rate significantly affects the speed and stability of the algorithm's convergence. An excessively large learning rate may skip the optimal solution, causing significant oscillations in the model parameters on the error surface, which in turn leads to severe fluctuations in the convergence curve and even potential divergence in the training process. Conversely, a learning rate that is too small results in slow parameter updates, significantly slowing down the convergence speed. Therefore, selecting an appropriate learning rate is crucial for the stable and efficient implementation of the algorithm.
\begin{figure}[t]
        \centering
        \includegraphics[width=8cm]{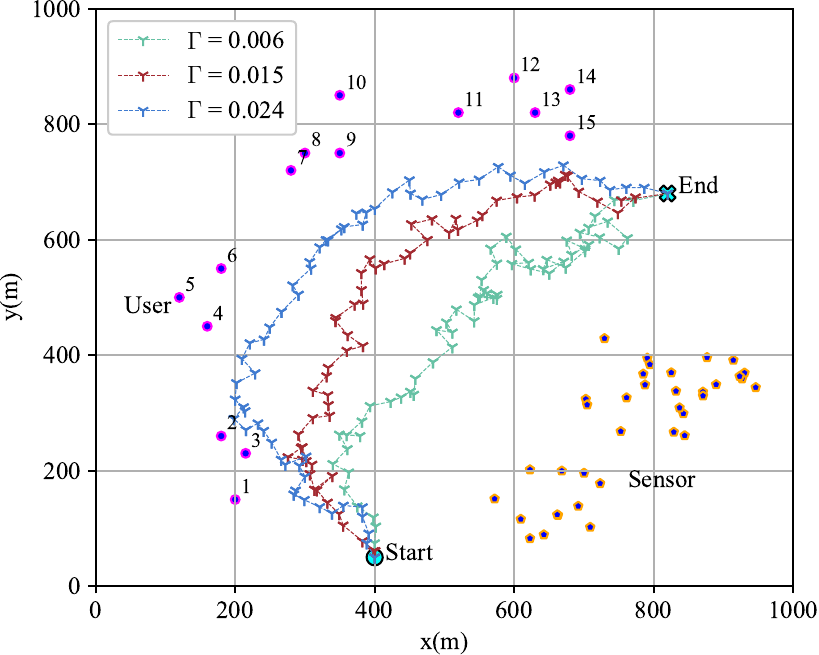}
        \vspace{-1.5mm}
        \caption{An illustration of UAV trajectories in separated topology (Topo.1).}
        \vspace{-3.5mm}
        \label{fig:topo1}
\end{figure}
\begin{figure}[t] 
  \centering
  \includegraphics[width=8cm]{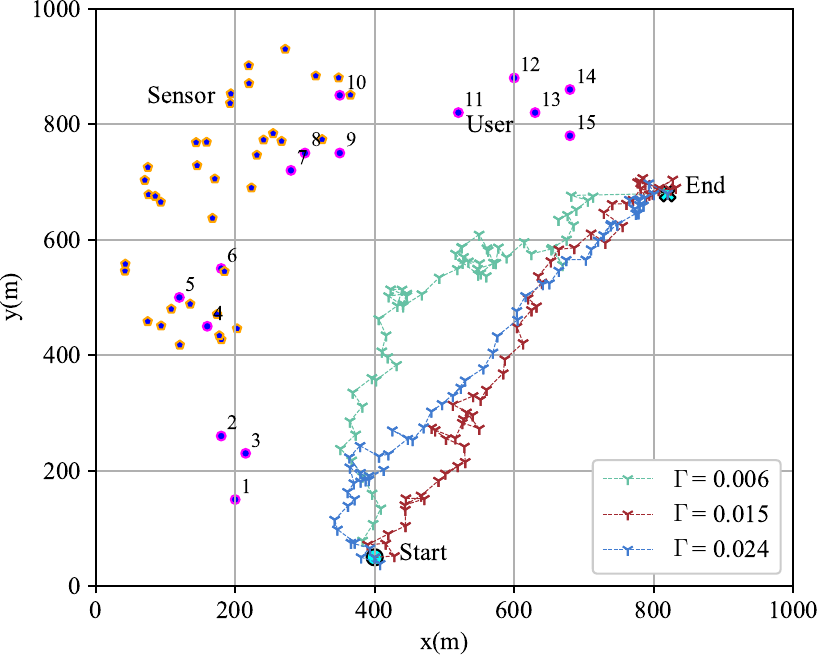}
  \vspace{-1.5mm}
  \caption{An illustration of UAV trajectories in mixed topology (Topo.2).}
  \vspace{-2.5mm}
  \label{fig:topo2}
\end{figure}
\begin{figure}[t] 
	\centering
	\includegraphics[width=7.5cm]{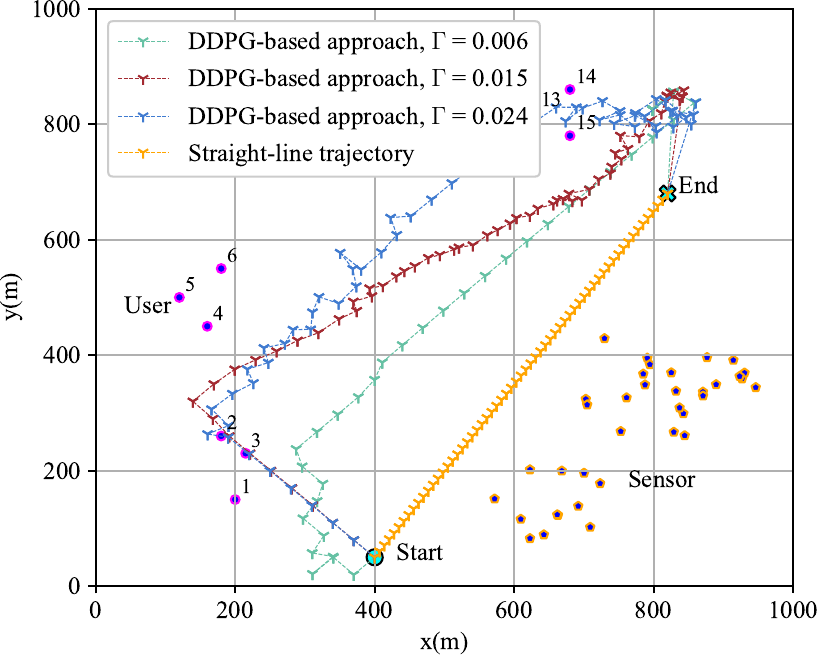}
	\vspace{-1.5mm}
	\caption{An illustration of UAV trajectories of comparison schemes.}
	\vspace{-3.5mm}
	\label{fig:topo_ddpg}
\end{figure}
\begin{figure}[t] 
        \centering
        \includegraphics[width=7.5cm]{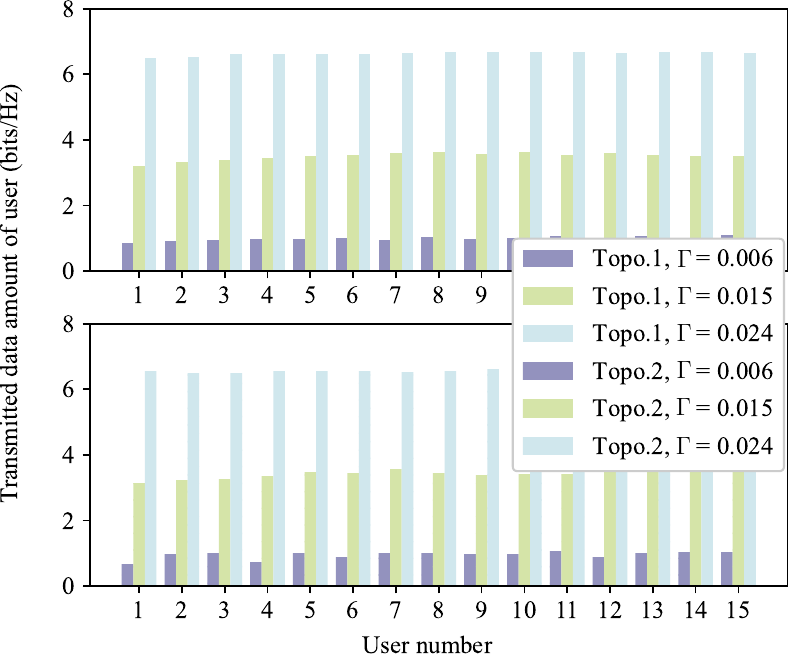}
        \vspace{-1.5mm}
        \caption{The total amount of data transmitted by users versus MSE threshold.}
        \vspace{-3.5mm}
        \label{fig:data}
\end{figure}
\begin{figure}[t]
	\centering
	\includegraphics[width=7.5cm]{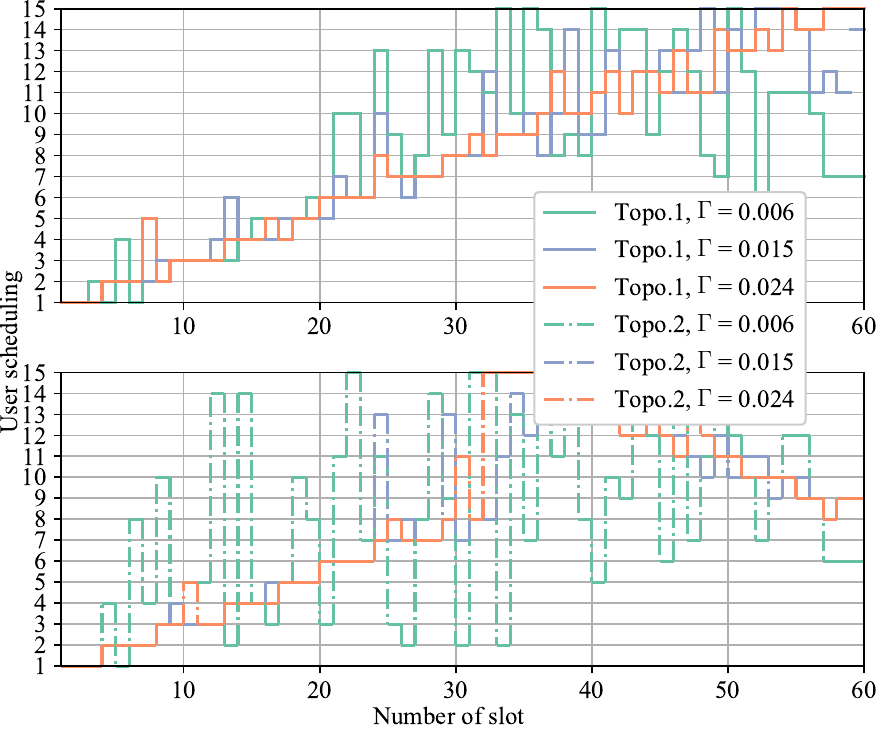}
	\vspace{-3.5mm}
	\caption{User scheduling scheme over the timeline.}
	\vspace{-3.5mm}
	\label{fig:access}
\end{figure}

In Figs.~\ref{fig:topo1} and~\ref{fig:topo2}, we compare the UAV trajectories optimized by the SAC-based algorithm under different AirComp accuracy constraints in both separated and mixed distribution topologies. The separated topology (Topo.1) refers to scenarios where the distribution areas of users and sensors are far apart, while the mixed topology (Topo.2) indicates that the distribution areas are close. It is evident that the UAV trajectories are significantly influenced by the computational accuracy requirements of AirComp, and the trend of UAV trajectory changes varies in different topologies. Specifically, in the dispersed topology shown in Fig.~\ref{fig:topo1}, where the positions of users and sensor devices are far apart, the higher the accuracy requirement of AirComp, the more the UAV trajectory tends to approach the sensors to provide better communication quality and meet the computational accuracy requirements. Conversely, lower computational accuracy requirements cause the UAV trajectory to lean more toward the users to achieve higher transmission rates. In the mixed topology shown in Fig.~\ref{fig:topo2}, where the positions of users and sensor devices are closer together, a similar trend is observed. When the accuracy requirement of AirComp is higher, the UAV trajectory tends to approach the sensors. When the accuracy requirement is lower, the UAV trajectory tends to maintain a certain distance from the sensors to avoid AirComp interference with user transmission, which could degrade system performance. Therefore, it is evident that optimizing UAV trajectory requires balancing user transmission performance and AirComp accuracy. In addition, Fig.~\ref{fig:topo_ddpg} shows the UAV trajectories of the straight-line trajectory and the DDPG-based algorithm under different MSE thresholds. It can be observed that the UAV trajectories based on the DDPG algorithm are similar to those of the proposed SAC-based algorithm, while the fixed trajectory of the baseline explains its lower performance.
\vspace{-1.5mm}
\subsection{Performance Comparison}
In Fig.~\ref{fig:data}, we present the performance of the proposed scheme under different AirComp computational accuracy requirements in different topologies. To provide a more intuitive demonstration, we use the total amount of data transmitted by users as the performance metric instead of the transmission rate. It can be observed that the total data transmitted by users increases with the MSE threshold. This is expected, as a higher MSE threshold implies more lenient AirComp accuracy requirements, allowing resources such as UAV trajectory and transmission power to be more effectively utilized to enhance user transmission. Additionally, under the same MSE threshold constraint, there is no significant difference in the total data transmitted by each user, reflecting the fairness of the user scheduling strategy. Then, by combining the user scheduling timeline in Fig.~\ref{fig:access} with the UAV trajectory and user indices in Figs.~\ref{fig:topo1} and~\ref{fig:topo2}, we can see that the UAV tends to access users closer to it first. This is because, generally, users closer to the UAV have better data transmission conditions, making it easier to achieve higher system performance. This further validates the effectiveness and stability of the proposed scheme.
\begin{figure}[t]
	\centering
	\includegraphics[width=7.5cm]{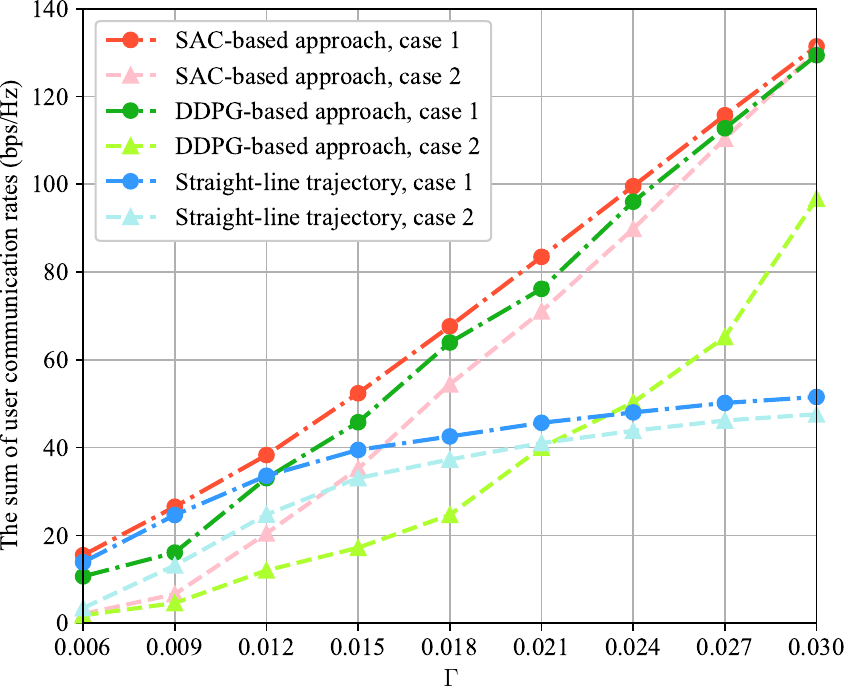}
	\vspace{-1.5mm}
	\caption{Performance comparison with different approaches for outer problem.}
	\vspace{-4.5mm}
	\label{fig:alg}
\end{figure}
\begin{figure}[t]
	\centering
	\includegraphics[width=7.5cm]{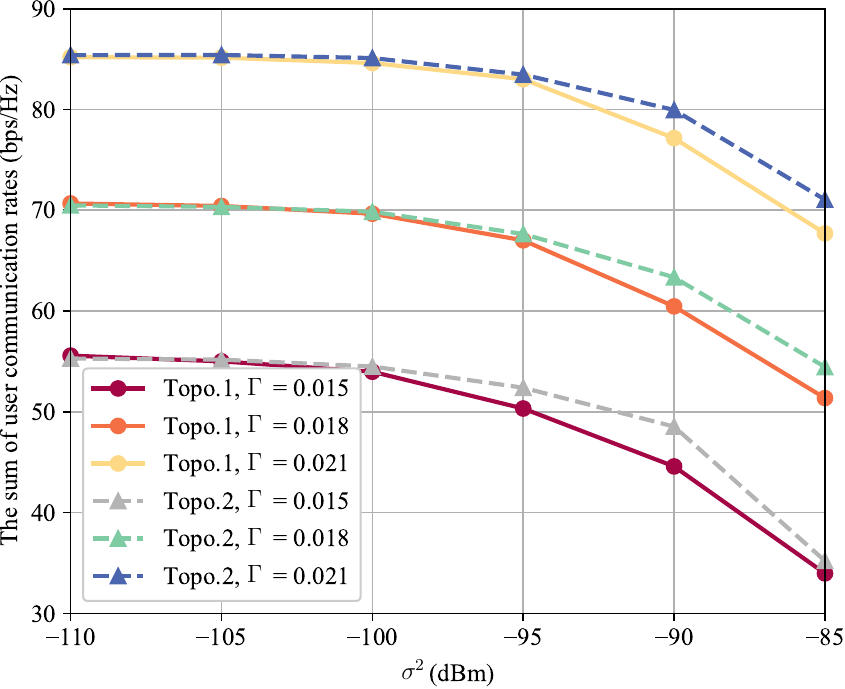}
	\vspace{-1.5mm}
	\caption{The performance of proposed SAC-based approach versus noise.}
	\vspace{-3mm}
	\label{fig:sigma}
\end{figure}

In Fig.~\ref{fig:alg}, we compare the performance of the proposed SAC-based approach, the DDPG-based approach, and the baseline with a straight-line trajectory and nearest user access. Notably, we focus on comparing the system performance under different schemes of UAV trajectory and user scheduling to highlight the performance differences in optimizing external problems. Similar to the results in Fig.~\ref{fig:access}, the system performance for all considered methods increases with the MSE threshold due to the trade-off between user transmission and AirComp accuracy. The proposed SAC-based algorithm provides the best performance, while the DDPG-based method is slightly inferior, and the baseline shows a significant performance gap. The performance of the proposed SAC-based algorithm can be further verified through comparison with a global search method, indicating that it achieves performance close to the global optimum. Furthermore, when the noise power increases, our proposed method performs better. This is because the SAC algorithm is inherently more stable and robust than the DDPG algorithm, allowing it to better adapt to environmental changes.
\begin{figure}[t]
	\centering
	\includegraphics[width=8cm]{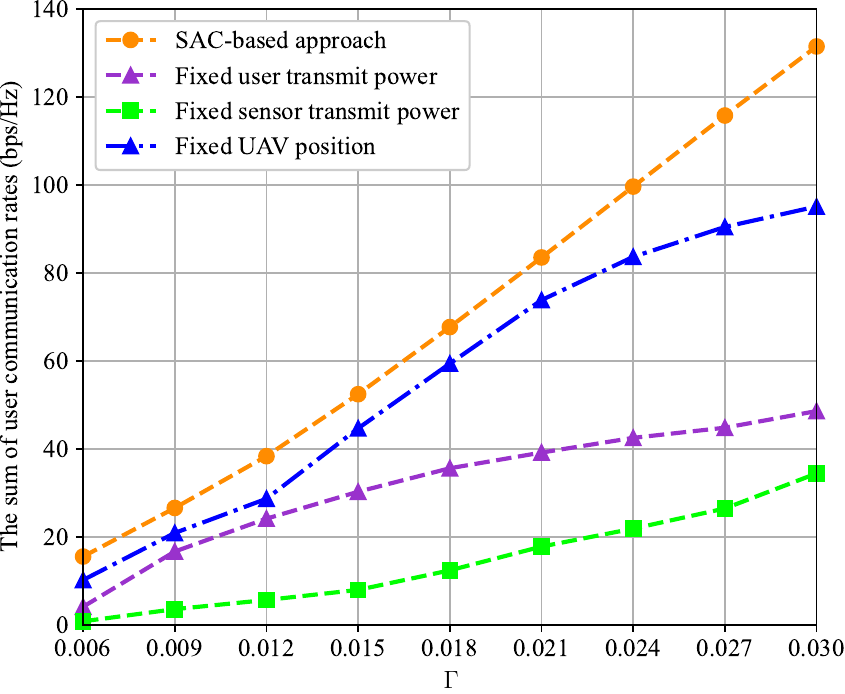}
	\vspace{-1.5mm}
	\caption{Performance comparison with different strategies versus MSE threshold.}
	\vspace{-4.5mm}
	\label{fig:diffing}
\end{figure}
\begin{figure}[t]
	\centering
	\includegraphics[width=7.5cm]{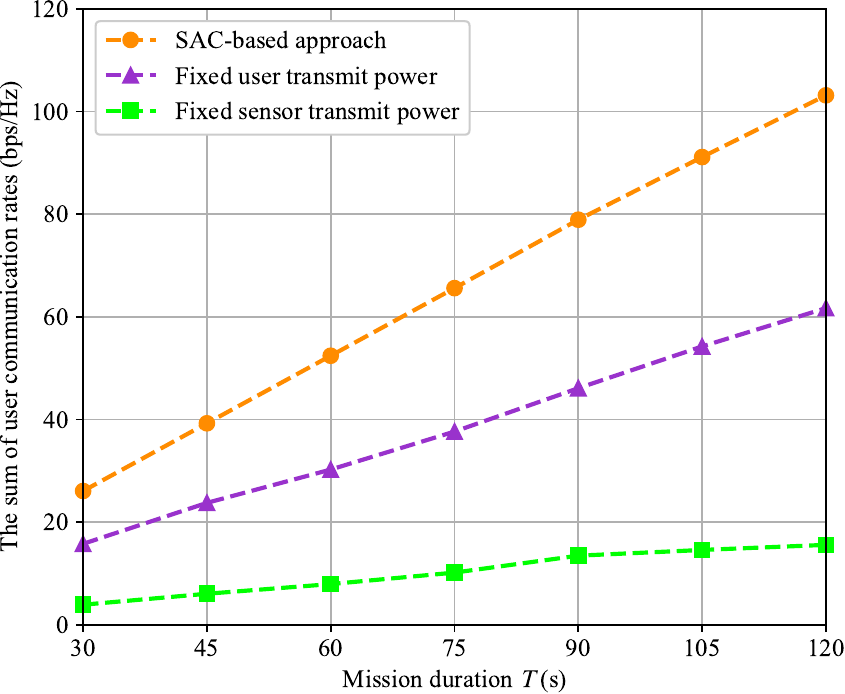}
	\vspace{-1.5mm}
	\caption{Performance comparison with different strategies versus mission time.}
	\vspace{-5mm}
	\label{fig:diffint}
\end{figure}

Fig.~\ref{fig:sigma} shows the performance under our proposal with the cases with different noise powers, where case 1 has a noise power of -95~dBm and case 2 has a noise power of -85~dBm. It can be observed that performance decreases with increasing noise power, regardless of whether it is in a dispersed topology, a mixed topology, or under different MSE threshold constraints. This is because, as noise increases, the air-to-ground channel conditions gradually deteriorate, impacting user transmissions and AirComp performance. This trend becomes more pronounced when noise exceeds -95~dBm. Meanwhile, the performance improves as the MSE threshold increases because looser accuracy requirements for AirComp result in more resources being available for communication.

In Fig.~\ref{fig:diffing}, we show the performance differences among the proposed SAC-based approach, fixed UAV position scheme, fixed user transmission power scheme, and fixed sensor transmission power scheme. The latter two schemes are defined as all users and sensors using the fixed transmit power, respectively, with the rest being identical to the proposed SAC-based method. As expected, the proposed SAC-based method achieves the best performance, while there is an evident performance gap with the comparison schemes. This demonstrates the effectiveness and importance of optimizing user transmission power and sensor transmission coefficients. Furthermore, as the MSE threshold increases, the performance of the fixed sensor transmission power scheme tends to stabilize. This is because the fixed user transmission strategy limits the system's performance, even though more resources, such as UAV trajectory and normalization factor, are allocated to communication. Fig.~\ref{fig:diffint} shows the performance differences between the proposed SAC-based method and the comparison schemes under different mission durations, demonstrating that our proposal significantly outperforms the two comparison schemes. As the mission duration increases, the waypoints of the UAV can be more flexible, and the number of communications for each user also increases. Therefore, the sum of the transmission rates is also greater.
\vspace{-1.5mm}
\section{Conclusion and Future Work} \label{sec:con}
In this paper, we innovatively proposed a UAV-assisted integrated communication and AirComp with interference awareness, where the transmission power of users, transmit coefficients of sensors, normalizing factor, and UAV trajectory were jointly considered. The proposed SAC-based method was exploited to ensure the accuracy requirements of AirComp, alleviate interference, and maximize the quality of transmission for users. We particularly proposed the reward function incorporates the UAV destination requirement and the fairness of user scheduling. The numerical results indicated that implicit coordination is required for the accuracy of AirComp and the user transmission rate in terms of UAV trajectory planning for the optimized system performance. Meanwhile, The proposed system considers interference between communication and AirComp, offering a reference for practical applications with varying AirComp accuracy requirements. Moreover, our proposal not only has the superiority in performance but also has a certain degree of robustness as compared with the baselines.

In future work, building upon the proposed UAV-assisted integrated communication and AirComp framework, we intend to incorporate existing channel reconfiguration techniques, such as RIS and extra-large MIMO, to achieve efficient beamforming and interference suppression. This approach will selectively enhance channel quality tailored to diverse user requirements, which in turn strengthens interference management capabilities and network robustness~\cite{ris1, ris2, xlmimo}.
\vspace{-7mm}
\bibliographystyle{IEEEtran}
\bibliography{mainBib}

\end{document}